\documentclass[11pt]{article}

\usepackage{amsmath}
\usepackage{amssymb}
\usepackage[paper=a4paper,text={17cm,25cm},centering]{geometry}
\usepackage{graphicx}
\usepackage{rotating}
\usepackage{enumerate}
\usepackage{caption}

\def\be{\begin{equation}}
\def\ee{\end{equation}}

\def\vec#1{\underline{#1}}
\def\mod#1{|#1|}

\begin{document}

\title{Nuclear scattering configurations of onia in different frames}
\author{{Anh Dung Le${}^{(1)}$, Alfred H. Mueller${}^{(2)}$,
  St\'ephane Munier${}^{(1)}$}\\
  {\footnotesize\it  (1) CPHT, CNRS, \'Ecole polytechnique, IP Paris,
    F-91128 Palaiseau, France}\\
  {\footnotesize\it  (2) Department of Physics, Columbia University,
  New York, NY 10027, USA}}
\date{March 24, 2021}

\maketitle

\begin{abstract}
  In the scattering of a small onium off a large nucleus at high center-of-mass energies, when the parameters are set in such a way that the cross section at fixed impact parameter is small, events are triggered by rare partonic fluctuations of the onium, which are very deformed with respect to typical configurations. Using the color dipole picture of high-energy interactions in quantum chromodynamics, in which the quantum states of the onium are represented by sets of dipoles generated by a branching process, we describe the typical scattering configurations as seen from different reference frames, from the restframe of the nucleus to frames in which the rapidity is shared between the projectile onium and the nucleus. We show that taking advantage of the freedom to select a frame in the latter class makes possible to derive complete asymptotic expressions for some boost-invariant quantities, beyond the total cross section, from a procedure which leverages the limited available knowledge on the properties of the solutions to the Balitsky-Kovchegov equation that governs the rapidity-dependence of total cross sections. We obtain in this way an analytic expression for the rapidity-distribution of the first branching of the slowest parent dipole of the set of those which scatter. This distribution provides an estimator of the correlations of the interacting dipoles, and is also known to be related to the rapidity-gap distribution in diffractive dissociation, an observable measurable at a future electron-ion collider. Furthermore, our result may be formulated as a more general conjecture, that we expect to hold true for any one-dimensional branching random walk model, on the branching time of the most recent common ancestor of all the particles that end up to the right of a given position.
\end{abstract}


\setcounter{tocdepth}{2}
\tableofcontents

\section{Introduction}

Onium-nucleus scattering is an outstanding process to understand
theoretically, first because it is the simplest interaction process between
a (model) hadron and a nucleus, and second for its potential phenomenological
applications. Indeed, if the center-of-mass energy is sufficiently large,
this process can easily be factorized from deep-inelastic
electron-nucleus scattering cross sections~\cite{Nikolaev:1990ja,Ewerz:2006vd}.
The latter will be measured
at the future electron-ion collider (EIC)~\cite{Accardi:2012qut} which will be built
at the Brookhaven National Laboratory within the next decade,
and at still higher-energy proposed DIS experiments, such as the
Large Hadron-Electron Collider (LHeC) at CERN and the
Future Circular Collider (FCC) in electron-hadron mode
(FCC-eh)~\cite{Agostini:2020fmq}.
On the other hand, in proton-nucleus collisions, it turns out that
an appropriate Fourier transform of the onium-nucleus total
cross section is mathematically identical to the differential cross
section for producing a semi-hard jet of given transverse
momentum~\cite{Kovchegov:2001sc},
at least at next-to-leading logarithmic accuracy~\cite{Mueller:2012bn}.
An onium may also be a good starting point to
model dilute systems, such as heavy
mesons, or maybe even specific states of protons,
in order to understand theoretically some of their
universal properties.\footnote{%
  For a review of scattering in quantum chromodynamics in the regime of high energies,
  see e.g. Ref.~\cite{Kovchegov:2012mbw}.}

The center-of-mass energy (or, equivalently, total rapidity)
dependence of onium-nucleus forward elastic scattering amplitudes
is encoded in the Balitsky-Kovchegov (BK) evolution
equation~\cite{Balitsky:1995ub,Kovchegov:1999ua} established in the
framework of quantum chromodynamics (QCD).
When restricted to the relevant regime for the calculation
of total scattering cross sections involving a small onium,
the latter belongs to the wide universality class of non-linear
diffusion equations, the main representant of which
is the well-known Fisher~\cite{f} and Kolmogorov-Petrovsky-Piscounov~\cite{kpp}
(FKPP) equation. (For background on the FKPP equation, see for example
the reports~\cite{VANSAARLOOS200329,brunet:tel-01417420}; For a review on how
the FKPP equation appears in QCD, see e.g.~\cite{Munier:2009pc}).
This fact can be understood quite simply from a physical point of
view. On one hand, the forward scattering amplitude, the evolution
of which is described by the BK equation,
is tantamount to the probability that at least one gluon in the Fock state
of the onium at the time of the interaction, produced
by a cascade of gluon branchings, is absorbed by the nucleus.
On the other hand, with an appropriate initial condition,
the solution to the FKPP equation is the
probability that there is at least one particle
generated by a one-dimensional branching-diffusion process in
space which has a position larger than some predefined number.
At the level of the evolution equations, the asymptotic
equivalence between the BK and the FKPP equations
becomes manifest after the identification of the time variable in
the latter with the rapidity variable in the former,
of the space variable with (the logarithm of) the
transverse dipole size, and after taking the appropriate limit
of the BK equation~\cite{Munier:2003vc}.
At the level of the underlying stochastic processes,
the QCD evolution towards very high energies
is a gluon branching process which, when
the large-number-of-color limit is taken, boils down to the iteration
of independent one-to-two color dipole splittings~\cite{Mueller:1993rr}.
This process results in realizations of a specific branching random walk.

Other observables, such as diffractive
cross sections, can be formulated with the help of a system of
BK equations, as was first shown by
Kovchegov and Levin~\cite{Kovchegov:1999ji}.

In this paper, we shall analyze the scattering cross section per impact
parameter for onium-nucleus
collisions in the region in which it is much smaller than
unity, namely when the size of
the onium is very small compared to the saturation radius, 
and more specifically in the so-called ``scaling region'',
which is a well-known parametric region in which
the cross section does not depend on
the rapidity and on the size of the onium independently, but
through a scaling variable function of the latter
two~\cite{Stasto:2000er}. In particular, we shall study the interpretation
of that cross section in the framework
of the parton model, in terms of fluctuations of the partonic
content of the onium, in different reference frames related to each
other through longitudinal boosts.

Our motivation is twofold. First, boost invariance is a fundamental
symmetry of scattering amplitudes, and it is interesting to understand
theoretically how it is realized microscopically in this particular
regime of QCD, in which the interacting objects may be thought of as
sets of independent partons generated by a branching process.
Second, it is already well-known that
using boost-invariance of the scattering amplitudes\footnote{
  Of course, generally speaking, the existence of a symmetry implies constraints
  on observables and on theories.
  Interestingly enough, in the context of (toy models for) high-energy scattering,
  boost invariance led to stringent constraints on the form of the elementary
  processes (such as parton recombination, or other nonlinear processes that slow down
  parton evolution in the very high-density regime)
  which should be taken into account at ultra-high energies~\cite{Blaizot:2006wp}.
}
helps to formulate
the calculation of observables. For example,
the simplest proof of the BK equation consists in writing down the
change of the partonic content of the onium in an infinitesimal
boost, starting from the restframe of the onium.
Here, we will take advantage of boost invariance to
select a specific class of frames in which we will
be able to evaluate a particular probability distribution, which is a priori
very difficult to calculate, with the help of the limited
available knowledge of the solution to the BK equation.

The main outcome of our work is a partonic picture of the
scattering in different
frames, which turns out to enable the derivation of
an expression of the asymptotics of the probability
distribution of the rapidity at which the slowest ancestor
of all dipoles that interact with the nucleus has branched.
The latter quantity characterizes the correlations of the interacting dipoles.
While it is not directly an observable, it was shown to be related
to the rapidity gap distribution in diffractive dissociation
events~\cite{Mueller:2018zwx,Mueller:2018ned}. Last but not least,
it is a quantity of more general interest in the study of branching
random walks.

We shall start (Sec.~\ref{sec:formulations})
by formulating the scattering amplitude as well
as the distribution of the branching rapidity in two different ways:
A formulation (in terms of evolution equations) that can be implemented numerically,
and a formulation that will
set the basis for an approximation scheme exposed in
Sec.~\ref{sec:phenomodel} and used to arrive at analytical asymptotic
expressions. In Sec.~\ref{sec:numerics}, we compare
our analytical predictions to numerical solutions to the
complete equations, and we present our conclusions and some prospects
in Sec.~\ref{sec:conclusion}.
Appendix~\ref{app:integral} outlines the evaluation of a useful integral, and
Appendix~\ref{app:num} presents an alternative numerical model.


\section{Two formulations for the amplitudes\label{sec:formulations}}

In this paper, we shall address the following quantities:
\begin{itemize}
\item The forward elastic scattering amplitude $T_1$
  of the onium off the nucleus at a fixed impact parameter,
  or equivalently, the corresponding $S$-matrix element
  $S\equiv 1-T_1$;
\item The probability $T_2$ that at least two dipoles present in the
  Fock state of the onium in the considered frame at the time
  of the interaction are involved in the scattering, as well as
  a particular differential $G$:
  The probability distribution of the rapidity relative to the nucleus
  at which the slowest common ancestor of all interacting dipoles has branched.
\end{itemize}

The most straightforward formulation of the calculation
of $T_1$, $T_2$ and $G$ consists
in writing down evolution equations with respect to the total
rapidity~(Sec.~\ref{ssec:exact}).
We shall then introduce frame-dependent representations
of the solutions to such equations (Sec.~\ref{ssec:frame-dependent}).


\subsection{\label{ssec:exact}Exact evolution equations in the dipole model}

\subsubsection{\label{sssec:BK}Forward elastic amplitude and the Balitsky-Kovchegov equation}

The $S$-matrix element for the forward elastic interaction of a color dipole
of transverse (two-dimensional) size $\vec{r}$ with a nucleus at relative rapidity $Y$
obeys the Balitsky-Kovchegov (BK)
equation~\cite{Balitsky:1995ub,Kovchegov:1999ua}
\be
\partial_Y S(Y,r)=\bar\alpha\int\frac{d^2 \vec{r}'}{2\pi}\frac{r^2}{r'^2(\vec{r}-\vec{r}')^2}
\left[S(Y,r')S(Y,\mod{\vec{r}-\vec{r}'})-S(Y,r)\right],
\label{eq:BK0}
\ee
where we have assumed homogeneity and isotropy
($S$ only depends on the modulus $r$ of $\vec{r}$, not on its orientation
nor on the absolute position of the dipole in the transverse plane):
In practice, this holds for nearly central
collisions of small dipoles with very extended nuclei.
Furthermore, this very equation is the lowest-order approximation of the QCD
evolution\footnote{%
  In the considered limit, we keep only the largest terms in the perturbative
  expansion of~$S$ when $Y$ is large and $\bar\alpha Y\sim 1$,
  which turn out to be the set of powers of $\bar\alpha Y$.
} in the limit of large~$Y$, large atomic number,
large number of colors $N_c$.
The constant $\bar\alpha$ that controls the pace of the evolution reads
$\bar\alpha\equiv\alpha_s N_c/\pi$, $\alpha_s$ being the QCD coupling.

The simplest way to derive this evolution equation is to start
from the restframe of the onium in which the nucleus is evolved
at rapidity $Y$, and to interpret $S(Y,r)$ as the probability
that an onium of size $r$, in its bare state, does not
interact with the nucleus. Then, one increases the total scattering rapidity
boosting the onium by $dY$, keeping the rapidity of
the nucleus fixed. In the QCD dipole model~\cite{Mueller:1993rr},
the scattering configuration of the initial onium may then either
become a set of two color dipoles of sizes $\vec{r}'$ and $\vec{r}-\vec{r}'$
(up to $d^2\vec{r}'$), with probability
\be
\bar\alpha\,dY\,dp_{1\rightarrow 2}(\vec{r},\vec{r}')\equiv
\bar\alpha\,dY\frac{d^2 \vec{r}'}{2\pi}\frac{\vec{r}^2}{{r'}^2(\vec{r}-\vec{r}')^2},
\ee
or may stay a single dipole, with probability\footnote{%
  An ultraviolet cutoff is understood in all integrations over the
  dipole sizes, which can eventually be set to zero in the
  equations for the physical observables.}
$1-\bar\alpha \,dY\int_{\vec{r}'}dp_{1\rightarrow 2}(\vec{r},\vec{r}')$.
Hence
\be
S(Y+dY,r)=\left(1-\bar\alpha\,dY\int_{\vec{r}'}dp_{1\rightarrow 2}(\vec{r},\vec{r}')\right)S(Y,r)
+\bar\alpha\,dY\int_{\vec{r}'}dp_{1\rightarrow 2}(\vec{r},\vec{r}')S(Y,r')S(Y,\mod{\vec{r}-\vec{r}'}),
\ee
from which Eq.~(\ref{eq:BK0}) easily follows.

The constant $\bar\alpha$ always enters as a scaling factor of the rapidity:
Therefore, it is convenient to absorb it into the rapidity
variable, defining $y\equiv\bar\alpha Y$.
From now on, we will exclusively use
this rescaled rapidity, which we will nevertheless
keep calling ``rapidity''.
With the help of these notations, the BK equation reads
\be
\partial_y S(y,r)=\int_{\vec{r}'}dp_{1\rightarrow 2}(\vec{r},\vec{r}')
\left[S(y,r')S(y,\mod{\vec{r}-\vec{r}'})-S(y,r)\right].
\label{eq:BK}
\ee

The initial condition at rapidity $y=0$ corresponds to
the scattering amplitude
of the dipole with an unevolved nucleus: It is usually assumed to
have the McLerran-Venugopalan form~\cite{McLerran:1993ni}
\be
S(y=0,r)=
\exp\left[{-\frac{r^2Q_A^2}{4}
    \ln\left(e+\frac{1}{r^2\Lambda_{\text{QCD}}^2}\right)}\right],
\label{eq:MV}
\ee
where the momentum $Q_A$, called the ``saturation momentum'',
is characteristic of the nucleus. (Its value
is of the order of 1~GeV for a large nucleus). In this model, the amplitude
$T_1\equiv 1-S$ is steeply falling from~1 to~0 as $r$ becomes smaller,
especially since the relevant scale for the dipole sizes
is logarithmic: As a matter of fact, it is a Gaussian function of this
variable. The typical value of $r$ at which the transition between
$S=0$ and $S=1$ happens is $r\sim 1/Q_A$.
The function $S$ is almost
tantamount to a Heaviside distribution
\be
S(y=0,r)\simeq \Theta(-\ln \,r^2Q_A^2).
\label{eq:Heaviside}
\ee

The solution to the BK equation is known asymptotically
\cite{Mueller:2002zm,Munier:2003vc}:\footnote{%
  Some subleading corrections to Eq.~(\ref{eq:T1BK}) are also known, see
  e.g. Ref.~\cite{Enberg:2005cb}, and Ref.~\cite{Albacete:2004gw} for an
  extensive numerical study; but we will not need them in the present work.
  }
\be
T_1(y,r)=1-S(y,r)\simeq
\text{const}\times\ln\frac{1}{r^2Q_s^2(y)}
\left[rQ_s(y)\right]^{2\gamma_0}
\exp\left(
-\frac{\ln^2[r^2Q_s^2(y)]}{2\chi''(\gamma_0)y}
\right),
\label{eq:T1BK}
\ee
where $\chi(\gamma)\equiv 2\psi(1)-\psi(\gamma)-\psi(1-\gamma)$, $\gamma_0$
solves $\chi'(\gamma_0)=\chi(\gamma_0)/\gamma_0$ and
\be
\ln \frac{Q_s^2(y)}{Q_A^2}=
\chi'(\gamma_0) y-\frac{3}{2\gamma_0}\ln y,
\label{eq:satscale}
\ee
up to an additive constant of order unity, which, in the limits of interest here,
may always be absorbed into a rescaling of the overall constant in $T_1$.
This expression is correct for very large values of $y$. In particular,
the logarithmic singularity for  $y\rightarrow 0$ would be regularized after
resummation of higher orders,
in such a way that $\ln({Q_s^2(y)}/Q_A^2)\underset{y\rightarrow 0}{\longrightarrow}0$.
$Q_s(y)$ is the saturation momentum of the nucleus at rapidity $y$,
namely $1/Q_s(y)$ is the typical value of the transverse size of a dipole that
interacts with it at the rapidity above which the dipole gets absorbed with
probability of order unity.
Equation~(\ref{eq:T1BK}) is only valid for
$1<\ln^2 [r^2Q_s^2(y)]\lesssim\chi''(\gamma_0)y$, which, up to strong
inequalities, defines the scaling region.
The numerical values of the parameters $\gamma_0$, $\chi'(\gamma_0)$, $\chi''(\gamma_0)$
can be found e.g. in Ref.~\cite{Mueller:2002zm}.
They are not relevant in our discussions:
The only important point is that they are all of order~1.

Note that the BK equation is also the evolution equation for the probability
that there is no dipole larger than $1/Q_A$ in the state of the onium evolved
to the rapidity $y$, when the initial condition is taken to be exactly the Heaviside
distribution~(\ref{eq:Heaviside}).


\subsubsection{Multiple scatterings}

The set of dipoles which interact with the nucleus necessarily stem from
the branchings of a single dipole: their ``last common ancestor''.
This is because we start the evolution with a single dipole (the onium),
and because, as manifest in a Hamiltonian formulation of QCD, in the
absence of recombination mechanism, partons evolve with
rapidity through elementary $1\rightarrow 2$
splitting processes.\footnote{There
  is also a $1\rightarrow 3$ process at next-to-leading order,
  but it would not fundamentally change our discussion.}
We want to compute the
distribution of the rapidity $y_1$, with respect to the nucleus,
at which this ancestor has branched.

Let us call $G(y,r;y_1)dy_1$ the joint probability
that the onium of initial size $r$ interact with the nucleus
and that the splitting rapidity of the last common ancestor
be $y_1$ up to $dy_1$,
the total rapidity of the interaction being $y$.
An evolution equation may be obtained,
using the same method as for $S$.
One starts with the frame in which the nucleus is boosted
to the rapidity $y\geq y_1$, while the onium of size $r$ is at rest.
One then increases the total rapidity by $dy$, keeping $y_1$ fixed,
through an infinitesimal boost of the onium.
In this rapidity interval, the onium may split to two dipoles
with probability $dp_{1\rightarrow 2}(\vec{r},\vec{r}') dy$, or stay
a single dipole with probability $1-\int dp_{1\rightarrow 2}(\vec{r},\vec{r}') dy$.

For an ensemble of events restricted to those without branching,
$G(y+dy,r;y_1)$ is just $G(y,r;y_1)$.
For events in which, instead, the initial dipole branches,
one and only one of the offspring dipoles may scatter.
So in this case, $G(y,r;y_1)$ is replaced by two terms each
consisting in the product of a factor $G$ and a factor $S$,
the arguments of which are either $r'$ or $\mod{\vec{r}-\vec{r}'}$.
Taking the sum over all possible events weighted by their probabilities,
we get
\begin{multline}
  G(y+dy,r;y_1)=
  \left(1-dy\int_{\vec{r}'}dp_{1\rightarrow 2}(\vec{r},\vec{r}')\right)G(y,r;y_1)\\
+dy\int_{\vec{r}'}dp_{1\rightarrow 2}(\vec{r},\vec{r}')
\left[
  G(y,r';y_1)S(y,\mod{\vec{r}-\vec{r}'})
  +G(y,\mod{\vec{r}-\vec{r}'};y_1)S(y,r')
  \right].
\end{multline}
Enforcing the limit $dy\rightarrow 0$, we obtain the evolution in the form of
the following integro-differential equation:
\be
\partial_y G(y,r;y_1)=\int_{\vec{r}'}dp_{1\rightarrow 2}(\vec{r},\vec{r}')\left[
  G(y,r';y_1)S(y,\mod{\vec{r}-\vec{r}'})+G(y,\mod{\vec{r}-\vec{r}'};y_1)S(y,r')-G(y,r;y_1)
  \right].
\label{eq:evolG}
\ee
The initial condition has to be set
when the total rapidity coincides with the splitting rapidity
of the common ancestor: $y=y_1$. In this case, there is no
choice: The onium has to branch at this very rapidity $y_1$, and each
offspring must scatter. This translates into the following equation:
\be
G(y_1,r;y_1)=\int_{\vec{r}'}dp_{1\rightarrow 2}(\vec{r},\vec{r}')\left[1-S(y_1,r')\right]
\left[1-S(y_1,\mod{\vec{r}-\vec{r}'})\right].
\label{eq:initG}
\ee
The equations~(\ref{eq:evolG}),(\ref{eq:initG})
for $G$ were written for the first time
in Ref.~\cite{Dung:2018,Anh:2019dea}. They were compared to the Kovchegov-Levin equations
for the rapidity-gap distribution in diffractive
dissociation, and solved numerically.

Let us introduce the probability $T_2$ that there are at least
two scatterings with the nucleus boosted to the rapidity~$y_0$. This
is just an integral of $G$ over $y_1$:
\be
T_2(y,r;y_0)=\int_{y_0}^{y} dy_1\,G(y,r;y_1).
\ee
A similar quantity has recently been identified as an estimator
of the contribution of higher twists to
total cross sections~\cite{grossberndt2020second}.
(Note that evidence for higher-twist effects was also found earlier
in the DESY-HERA data
for diffractive deep-inelastic scattering, see e.g. Ref.~\cite{Motyka:2012ty}).

$T_2$ actually obeys an evolution equation straightforward to deduce from
the evolution equation for $G$, which we may write as
\begin{multline}
  \partial_y T_2(y,r;y_0)=
  \int_{\vec{r}'}dp_{1\rightarrow 2}(\vec{r},\vec{r}')
  \bigg[
  T_2(y,r';y_0)+T_2(y,\mod{\vec{r}-\vec{r}'};y_0)-T_2(y,r;y_0)\\
  -T_2(y,r';y_0)T_1(y,\mod{\vec{r}-\vec{r}'})
  -T_1(y,r')T_2(y,\mod{\vec{r}-\vec{r}'};y_0)
  +T_1(y,r')T_1(y,\mod{\vec{r}-\vec{r}'})
  \bigg],
  \label{eq:evolution-T2}
\end{multline}
with the initial condition at $y=y_0$ which reads $T_2(y_0,r;y_0)=0$.

The evolution equations (\ref{eq:evolG}) and (\ref{eq:evolution-T2})
for $G$ and $T_2$ respectively
may be solved numerically (with the help of a solution to
the BK equation~(\ref{eq:T1BK})),
but no analytical expression is known.
As we will see, we can however obtain exact asymptotic expressions
for these quantities (actually for the ratios $G/T_1$ and $T_2/T_1$) in a model
expected to capture the main features of the QCD dipole model and of
more general branching random walks.
The starting point will be a useful representation of $T_2$ in terms of
dipole densities and of the dipole-nucleus scattering amplitude $T_1$, that we shall
expose in the next section.


\subsection{\label{ssec:frame-dependent}Frame-dependent representations}

In this section, we discuss a representation of the solutions to
these evolution equations which will prove useful to set up
approximation schemes, from which we shall find asymptotic expressions.

Let us choose a frame in which the nucleus is boosted at rapidity $y_0$,
and the onium at rapidity $\tilde y_0\equiv y-y_0$ in the opposite sense.
We will consider frames defined by a large~$y_0$, but $\tilde y_0$
will not be smaller than a non-negligible fraction of the total rapidity~$y$.

Instead of using as variables the sizes $r$ of the dipoles or the saturation
momentum at rapidity $y$, $Q_s(y)$, we shall express
all functions with the help of the logarithms of these sizes and of the momentum,
defining
\be
x\equiv \ln \frac{1}{r^2Q_A^2}
\quad\text{and}\quad
X_y\equiv \ln \frac{Q_s^2(y)}{Q_A^2}.
\label{eq:notation_log-size}
\ee

\subsubsection{$S$-matrix element\label{sssec:S-matrix}}

The following formula is an exact representation of the solution
to Eq.~(\ref{eq:BK}):
\be
S(y,x)=\left\langle\prod_{\{x_i\}}S(y_0,x_i)\right\rangle_{\tilde y_0,x},
\label{eq:frame-dependent-1}
\ee
where the averaging is over all the dipole configurations
of the onium at rapidity $\tilde y_0$ (with respect to the onium),
represented by the set of log-sizes $\{x_i\}$.
The functions $S$ that appear left and right are the
same, but evaluated at two different rapidities. Because of
boost invariance, $S$ in the lefthand
side must be independent of $y_0$ chosen in the righthand side.

$S$ defined in Eq.~(\ref{eq:frame-dependent-1}) obeys the BK
equation~(\ref{eq:BK}).
To check this statement, it is enough to see that increasing $y$
by $dy$ amounts to increasing $\tilde y_0$ by the same $dy$, and
to decomposing the averaging over the dipole configurations at $\tilde y_0+dy$
as
\begin{multline}
\left\langle
\prod_{\{x_i\}}S(y_0,x_i)
\right\rangle_{\tilde y_0+dy,x= \ln\frac{1}{r^2Q_A^2}}
=\left(1-dy\int_{\vec{r}'}dp_{1\rightarrow 2}(\vec{r},\vec{r}')\right)
\left\langle\prod_{\{x_i\}}S(y_0,x_i)\right\rangle_{\tilde y_0,x=\ln \frac{1}{r^2Q_A^2}}\\
+dy\int_{\vec{r}'}dp_{1\rightarrow 2}(\vec{r},\vec{r}')
\left\langle\prod_{\{x'_i\}}S(y_0,x'_i)\right\rangle_{\tilde y_0,
  x'\equiv\ln\frac{1}{r^{\prime 2}Q_A^2}}
\left\langle\prod_{\{x''_i\}}S(y_0,x''_i)
\right\rangle_{\tilde y_0,x''\equiv \ln \frac{1}{(\vec{r}-\vec{r}^{\prime})^2Q_A^2}},
\end{multline}
where the sets $\{x'_i\}$ and $\{x''_i\}$ represent
the dipole configurations at rapidity $\tilde y_0$ of initial dipoles
of respective sizes $\vec{r}'$ and $\vec{r}-\vec{r}'$.
Simple manipulations and replacements lead to Eq.~(\ref{eq:BK}),
after the limit $dy\rightarrow 0$ has been taken.

We may rewrite Eq.~(\ref{eq:frame-dependent-1})
with the help of the number density $n(x)$ of dipoles
of log-size $x$:
\be
S(y,x)=\left\langle
\prod_{x'}\left[S(y_0,x')\right]^{n(x') dx'}
\right\rangle_{\tilde y_0,x},
\ee
where the product is now over all the bins in dipole size, of width $dx'$.
Note that $n(x')$ is a random density, the distribution of which
depends on the size of the initial dipole
and on the evolution rapidity $\tilde y_0$.
This equation can be expressed for $T_1$:
\be
T_1(y,x)=\left\langle
1-\exp\left\{\int dx'\,{n(x')}\ln\left[
  1-T_1(y_0,x')
  \right]
\right\}
\right\rangle_{\tilde y_0,x}.
\label{eq:T1bootstrap0}
\ee

Now, we assume that the dipoles that effectively contribute
to the integral all have log-sizes $x'$ such that $T_1(y_0,x')\ll 1$.
This is verified if the onium configurations which contain individual
dipoles larger than the inverse saturation scale of the nucleus only
bring a negligible contribution to the overall amplitude.
We will check a posteriori that it is a consistent assumption.
In the framework of this approximation, we can expand the logarithm in
Eq.~(\ref{eq:T1bootstrap0}), and deduce an elegant
formula for the amplitude $T_1=1-S$:
\be
T_1(y,x)=\left\langle
1-e^{-I(y_0)}
\right\rangle_{\tilde y_0,x},
\label{eq:T1bootstrap}
\ee
where we have introduced the notation
\be
I(y_0)=\int dx'\, n(x')\, T_1(y_0,x')
\ee
for the overlap of the dipole-nucleus scattering amplitude and
the dipole density in the onium.


\subsubsection{Contribution of multiple scatterings}

Let us compute the amplitude for scattering with at least
two exchanges between the configuration of the onium at rapidity $\tilde y_0$
and the nucleus evolved to the rapidity $y_0$. The exact formula reads
\be
T_2(y,x;y_0)=\left\langle
1-\left(1+\sum_{\{x_i\}}\frac{T_1(y_0,x_i)}{S(y_0,x_i)}\right)
\prod_{\{x_i\}}S(y_0,x_i)
\right\rangle_{\tilde y_0,x}.
\label{eq:frame-dependent-T2}
\ee
In the same way as in the case of the $S$-matrix element,
we can show that the righthand side of
Eq.~(\ref{eq:frame-dependent-T2})
obeys the evolution equation~(\ref{eq:evolution-T2}).

$T_2$ obviously depends on $y_0$.
$G$ instead, which is formally a rapidity derivative of $T_2$,
\be
G(y,r;y_1)=-\frac{\partial}{\partial y_0}\left.T_2(y,r;y_0)\right|_{y_0=y_1},
\label{eq:fromT2toG}
\ee
will be independent of the choice of frame.

Assuming again that $S(y_0,x_i)\simeq 1$ for all dipoles in
the relevant configurations, we get
\be
T_2(y,x;y_0)=\left\langle
1-\left[1+I(y_0)\right]
e^{-I(y_0)}
\right\rangle_{\tilde y_0,x}.
\label{eq:formulationT2}
\ee

We are now going to evaluate the right-hand sides of Eqs.~(\ref{eq:T1bootstrap})
and~(\ref{eq:formulationT2}). This cannot be done through a straightforward
calculation, but a simple model for the realizations of
branching random walks/dipole evolution can be used.


\section{\label{sec:phenomodel}Asymptotic amplitudes from
  the phenomenological model for front fluctuations}

In the following, we will stick to the large-rapidity limit, and
pick the size of the initial onium in the so-called scaling region.
This means that
\be
1\ll \ln^2 \frac{1}{r^2Q_s^2(y)}\ll y,
\quad\text{namely}\quad
1 \ll (x-X_y)^2 \ll y.
\label{eq:definition-scaling-region}
\ee
(A $\chi''(\gamma_0)$ factor would multiply $y$, but it is of order~1,
so it does not modify these strong inequalities.)
We shall actually take a slightly stronger condition on the lower bound on
$x-X_y$: We will always assume that it be much larger than the logarithms of the
rapidities $y$ and $y_0$.

Setting $r$ much smaller than $1/Q_s(y)$, as encoded in the
first strong inequality, implies that a typical realization of the dipole
evolution would interact with very small probability. So we need
fluctuations to create larger dipoles. We shall now introduce a model for these
fluctuations and apply it to the evaluation of $T_1$, $T_2$, $G$.

\subsection{Model for the dipole distribution\label{sec:model-dipole-dist}}

We present here a slightly modified formulation of the
model for the evolution of branching random
walks, and in particular for the QCD dipole evolution, that was
initially developed in Ref.~\cite{Mueller:2014gpa} and
applied to particle physics in Ref.~\cite{Mueller:2014fba}.

We assume that the evolution process develops essentially
in a deterministic, ``mean-field'' way, such that the density
of dipoles of log-size $x'$ at rapidity $y_i$ with
respect to the nucleus, namely after evolution over
the rapidity range $\tilde y_i$, reads
\be
\bar n(\tilde y_i,x'-x)=C_1(x'-x-\tilde X_{{\tilde y}_i})
e^{\gamma_0(x'-x-\tilde X_{\tilde y_i})}
\exp\left(
-\frac{(x'-x-\tilde X_{\tilde y_i})^2}{2\chi''(\gamma_0){\tilde y}_i}
\right)\Theta(x'-x-\tilde X_{\tilde y_i}),
\label{eq:barnpheno}
\ee
where
\be
\tilde X_{\tilde y_i}=-\chi'(\gamma_0){\tilde y_i}
+\frac{3}{2\gamma_0}\ln \tilde y_i.
\ee
There may be an additive constant of order~1 in $\tilde X_{\tilde y_i}$,
but to the accuracy we are
considering, it can be absorbed into the overall constant $C_1$.
Again, if this formula is to be extrapolated to the non-asymptotic regime
of $\tilde y_i$,
the logarithm has to be regularized in such a
way that $\tilde X_0=0$.

This formula represents the dipole density in a typical realization
of the dipole evolution, in the absence of a large fluctuation,
in a region of size of order $\sqrt{\tilde y_i}$ near the typical log-size
of the largest dipole, which is such that
$\ln\frac{1}{r_{\text{largest}}^2 Q_A^2}=x+\tilde X_{\tilde y_i}$.
In practice, it is obtained from the solution of a linearized BK (or FKPP)
equation with a cutoff that simulates the effect on the evolution of the
discreteness of the dipoles in realizations; see Ref.~\cite{Mueller:2014gpa}.

On top of this deterministic particle density,
we assume that one single fluctuation occurs after some random
evolution rapidity $\tilde y_1$, and that this fluctuation
consists in a dipole of size larger than the largest dipole in typical
configurations by a factor $e^{\delta/2}$. After this fluctuation has occurred,
the large produced dipole builds up into a second front in a deterministic
way upon further rapidity evolution.

We need the distribution of $\delta$. We
guess that it coincides with the probability of observing the
largest dipole with a log-size shifted by $(-\delta)$ with respect to the
mean-field tip of the distribution. This probability solves the BK equation
(see the remark at the end of Sec.~\ref{sssec:BK}),
and thus, has the same form as Eq.~(\ref{eq:T1BK}): The rate
at evolution rapidity $\tilde y_1$
reads, asymptotically for large $\tilde y_1$ and $\delta$,
\be
p(\delta,\tilde y_1)=C\,\delta\, e^{-\gamma_0\delta}
\exp\left(-\frac{\delta^2}{2\chi''(\gamma_0)\tilde y_1}\right)
\Theta(\delta).
\ee

In the kinematical region we consider, the mean-field evolution
of the initial onium
would not alone trigger a scattering: Hence the onium always scatters exclusively
through the smaller front that stems from the fluctuation.
Each dipole in the state of the onium at the interaction rapidity
scatters independently, with an amplitude $\bar T_1(y_0,x')$ that solves
the BK equation~(\ref{eq:BK}) with $S$ substituted with $1-\bar T_1$.
We shall denote by
\be
X_{y_0}=\chi'(\gamma_0)y_0-\frac{3}{2\gamma_0}\ln y_0
\ee
the log-saturation scale of the nucleus front at rapidity $y_0$,
see Eq.~(\ref{eq:satscale}) with the notation~(\ref{eq:notation_log-size}).

Let us now express $T_1$ and $T_2$ in this model.
The overlap of the amplitude $\bar T_1$
and of the dipole number density $\bar n$ that appear in
Eq.~(\ref{eq:T1bootstrap}) and~(\ref{eq:formulationT2}) reads,
in this model,
\be
I(y_0;\delta,y_1)\equiv \int dx'\, \bar n(\tilde y_0-\tilde y_1,x'
-\Xi_{\delta,\tilde y_1})\,
\bar T_1(y_0,x'),
\label{eq:Idef}
\ee
where $\Xi_{\delta,\tilde y_1}\equiv x+\tilde X_{\tilde y_1}-\delta$
is the log-size of the lead dipole at rapidity $\tilde y_1$. Then
\be
T_1(y,x)=\int_{y_0}^{y} dy_1\int_0^{\infty}d\delta\,
p(\delta,\tilde y_1)
\left(
  1-e^{-I(y_0;\delta,y_1)}
  \right),
\label{eq:T1pheno}
\ee
and
\be
T_2(y,x;y_0)=\int_{y_0}^{y} dy_1\int_0^{\infty}d\delta\,
p(\delta,\tilde y_1)
\left\{
  1-\left[1+I(y_0;\delta,y_1)\right]
e^{-I(y_0;\delta,y_1)}\right\}.
\label{eq:T2pheno}
\ee

We can also obtain a formula for $G$ itself in the framework of the
phenomenological model: It is enough to use Eq.~(\ref{eq:fromT2toG}),
from which one sees that the analytical expressions of
$T_2$ and $G$ only differ by the presence of the
integration over $y_1$ in the expression of the former, a fact which
is easy to understand. Indeed, the essence of the
phenomenological model is to single out one dipole in the state of
the onium evolved to the rapidity $\tilde y_1$ that will stand for the common
ancestor of all dipoles which scatter after evolution to
the rapidity $\tilde y_0$. When the $y_1$ integration in Eq.~(\ref{eq:T2pheno})
is left undone, then $G$ reads
\be
G(y,x;y_1)=
\int_0^{\infty}d\delta\,
p(\delta,\tilde y_1)
\left\{
  1-\left[1+I(y_0;\delta,y_1)\right]e^{-I(y_0;\delta,y_1)}\right\}.
\label{eq:Gpheno}
\ee

Let us introduce the distance, at the scattering rapidity,
between the tip of the dipole
distribution and the top of the nucleus front:
\be
\Delta(y_0;\delta,y_1)\equiv \tilde X_{\tilde y_0-\tilde y_1}+\Xi_{\delta,\tilde y_1}-X_{y_0}.
\label{eq:Delta_def}
\ee
In other words, $\Delta(y_0;\delta,y_1)$ is the logarithm of the squared ratio
of the size of the smallest dipole which would
scatter with probability of order unity with the nucleus in a scattering of
relative rapidity $y_0$, and of
the size of the largest dipole in the actual state of the onium
at rapidity $\tilde y_0$. It may be rewritten as
\be
\Delta(y_0;\delta,y_1)=x-X_y-\delta
+\frac{3}{2\gamma_0}\ln\frac{(\tilde y_0-\tilde y_1)y_0 \tilde y_1}{y}.
\label{eq:Delta}
\ee
As commented above (see e.g. after Eq.~(\ref{eq:barnpheno})), the logarithmic term
must be regularized in the limits $y_1\rightarrow y_0$ and $y_1\rightarrow y$. Furthermore,
with the considered choice of frame and parameters, this logarithmic term is always small
compared to $x-X_y$.

We shall first show that we may restrict ourselves to fluctuations such that $\Delta\geq 0$.
To this aim, we evaluate parametrically the contribution to $T_1(y,x)$
of the integration region
$\Delta\leq 0$, namely $\delta\geq\delta_{0}\equiv x-X_y
+\frac{3}{2\gamma_0}\ln\frac{(\tilde y_0-\tilde y_1)y_0 \tilde y_1}{y}$.
Starting from Eq.~(\ref{eq:T1pheno}), we see that we have the following upper bound
on the contribution of this region to $T_1$:
\be
\left.T_1(y,x)\right|_{\Delta\leq 0}
\leq\int_{y_0}^{y} dy_1\int_{\delta_{0}}^\infty d\delta\,
p(\delta,\tilde y_1).
\ee
The Gaussian factor in $p$ may be replaced by an
effective upper cutoff on the integration over $\delta$, set at
$\delta_{1}\equiv \sqrt{2\chi''(\gamma_0)\tilde y_1}$, and the integration
over $\delta$ can then be performed. The
condition $\delta_0\leq\delta_1$ for this integral not to be null implies that
$y_1\leq y-(x-X_y)^2/[2\chi''(\gamma_0)]$. All in all, we get a further bound on $T_1$:
\begin{multline}
\left.T_1(y,x)\right|_{\Delta\leq 0}\leq\frac{C}{\gamma_0}\frac{1}{y_0^{3/2}}
(x-X_y)e^{-\gamma_0(x-X_y)}\\
\times\int_{y_0}^{y-(x-X_y)^2/[2\chi''(\gamma_0)]} dy_1
\left.\left(
\frac{y}{(y-y_1)(y_1-y_0)}
\right)^{3/2}\right|_{\text{regularized}},
\end{multline}
where we have reminded that the apparent singularity at the lower bound of
the integral over $y_1$ needs to
be regularized, in such a way that the integrand remains finite of order~1.
The integral is now at most of order one.
Hence, we see that $T_1$ is suppressed by at least a factor $y_0^{3/2}\gg 1$ with
respect to the expected result, see Eq.~(\ref{eq:T1BK}). This proves that the
region $\Delta\leq 0$ can be neglected. From now on, we will only consider the
integration region in which $\Delta\geq 0$, i.e. $\delta\leq \delta_0$.

Actually, a closer look would show that only the region in which
$\Delta\gtrsim\frac{3}{2\gamma_0}\ln y_0$ contributes significantly.
This means physically
that the scattering amplitude of all the individual
dipoles in the fluctuation, at rapidity $\tilde y_0$,
is very small, consistently with the assumption that led
to Eqs.~(\ref{eq:T1pheno}),(\ref{eq:T2pheno}).

We see that all the functions $T_1$, $T_2$ and $G$ are written in terms of $p$ and $I$.
So let us express $I$ in the phenomenological model.
The function $\bar n$ that appears in Eq.~(\ref{eq:Idef})
is replaced by the expression given in Eq.~(\ref{eq:barnpheno}).
As for $\bar T_1$, since only the region $x'>X_{y_0}$ will be probed,
we can use the solution~(\ref{eq:T1BK}) of the BK equation
re-expressed in appropriate variables, namely
\be
\bar T_1(y_0,x')=C_2(x'-X_{y_0})e^{-\gamma_0(x'-X_{y_0})}
\exp\left(-\frac{(x'-X_{y_0})^2}{2\chi''(\gamma_0)y_0}\right)
\Theta(x'-X_{y_0}).
\label{eq:barT1pheno}
\ee
We get
\begin{multline}
I(y_0;\delta,y_1)=C_1C_2\,
e^{\gamma_0(X_{y_0}-\tilde X_{\tilde y_0-\tilde y_1}-\Xi_{\delta,\tilde y_1})}
\int dx'(x'-\tilde X_{\tilde y_0-\tilde y_1}-\Xi_{\delta,\tilde y_1})(x'-X_{y_0})\\
\times\exp\left[
  -\frac{(x'-\tilde X_{\tilde y_0-\tilde y_1}-\Xi_{\delta,\tilde y_1})^2}
  {2\chi''(\gamma_0)(\tilde y_0-\tilde y_1)}
  -\frac{(x'-X_{y_0})^2}{2\chi''(\gamma_0) y_0}\right]
\Theta(x'-\tilde X_{\tilde y_0-\tilde y_1}-\Xi_{\delta,\tilde y_1})\Theta(x'-X_{y_0}).
\end{multline}

We shall now compute the different quantities in the framework
of the phenomenological model.
$T_2$ is explicitly frame-dependent, and although $T_1$ and $G$ are boost-invariant,
their evaluation will depend upon the chosen frame.


\subsection{Amplitudes in a frame in which the nucleus
  is highly boosted}

In this section, we choose the frame in which the nucleus is boosted
to rapidity $y_0$ such that
\be
y_0\gg(x-X_y)^2\gg 1.
\ee

We shift the integration variable $x$, defining
$\bar x\equiv x'-\tilde X_{\tilde y_0-\tilde y_1}-\Xi_{\delta,\tilde y_1}$
the log-size of the dipoles
relative to the log-size of the largest dipole,
at the tip of the particle distribution.
The overlap integral then reads
\begin{multline}
I(y_0;\delta,y_1)=C_1 C_2\, e^{-\gamma_0\Delta(y_0;\delta,y_1)}\exp\left(
-\frac{\Delta^2(y_0;\delta,y_1)}{2\chi''(\gamma_0)y_1}
\right)\\
\times\int_{0}^{+\infty}d\bar x\,\bar x[\bar x+\Delta(y_0;\delta,y_1)]\,
\exp\left[
  -\frac{y_1}{2\chi''(\gamma_0) y_0 (\tilde y_0-\tilde y_1)}
  \left(\bar x+\frac{y_1-y_0}{y_1}\Delta(y_0;\delta,y_1)
  \right)^2
\right].
\end{multline}
We observe that the integral is determined by a large
integration region, up to $\bar x\sim \sqrt{y_0}$.
When $(x-\bar X_y)^2\ll y_0$,
we can neglect $\Delta(y_0;\delta,y_1)$ compared to $\bar x$, and the
integral takes a simpler form, which can be integrated exactly.
Moreover, we will check a posteriori that typically, $\tilde y_1\ll y$,
hence $y_1\sim y$, and the Gaussian factor involving $\Delta(y_0;\delta,y_1)$
can be set to unity.
Then we find
\be
I(y_0;\delta,y_1)=C_1 C_2\, e^{-\gamma_0\Delta(y_0;\delta,y_1)}
\frac{\sqrt{\pi}}{4}
\left(
  \frac{2\chi''(\gamma_0)y_0(\tilde y_0-\tilde y_1)}{y_1}
  \right)^{3/2}.
\ee
Replacing $\Delta(y_0;\delta,y_1)$ by its expression~(\ref{eq:Delta}),
we arrive at
\be
I(y_0;\delta,y_1)={
  C_1C_2 \sqrt{\frac{\pi}{2}} [{\chi''(\gamma_0)}]^{3/2}
e^{-\gamma_0(x-X_y)}}
\times\left(
\frac{y}{y_1\tilde y_1}
\right)^{3/2}
e^{\gamma_0\delta}.
\label{eq:I}
\ee
We see that $I$ turns out to be independent of $y_0$.
Let us introduce the following notation:
\be
p_1\equiv  C_1C_2 \sqrt{\frac{\pi}{2}} [{\chi''(\gamma_0)}]^{3/2}
e^{-\gamma_0(x-X_y)}.
\label{eq:def-p_1}
\ee
$p_1 e^{\gamma_0\delta}$ is just
the overlap of the front of the nucleus with that of an onium
if the latter were evolved in a purely deterministic way and starting
at a log-size $x-\delta$.


\subsubsection{Forward elastic scattering amplitude $T_1$\label{sssec:T1}}

The amplitude $T_1$ is obtained by substituting Eq.~(\ref{eq:I})
into Eq.~(\ref{eq:T1pheno}), with the restriction on $\delta$ imposed by the condition
$\Delta\geq 0$:
\be
T_1(y,x)=C\int_{y_0}^{y}dy_1
\int_0^{\delta_0} d\delta\,\delta
e^{-\gamma_0\delta}\exp\left(
-\frac{\delta^2}{2\chi''(\gamma_0)\tilde y_1}
\right)
\left\{
1-\exp\left[
  -p_1\left(
  \frac{y}{y_1\tilde y_1}
  \right)^{3/2}e^{\gamma_0\delta}
  \right]
  \right\}.
\ee
We shift $\delta$, defining the new integration variable
$\delta'\equiv\delta+\frac{3}{2\gamma_0}
\ln\frac{y}{y_1\tilde y_1}$.
Due to the $e^{-\gamma_0(x-X_y)}$ factor in $p_1$, the integration domain
extends effectively to
$\delta'\lesssim x-X_y$, a region much larger than the logarithm
of any rapidity appearing in this problem.
Hence, the lower integration bound on $\delta'$ may be kept to~$0$,
$\delta$ can just be identified to $\delta'$ in the Gaussian factor, and the upper
bound on $\delta$ can be released since the region $\delta>\delta_0\simeq x-X_y$ gets
anyway cutoff by the factors in the integrand.
We get
\be
T_1(y,x)=
C\int_0^\infty d\delta'\,\delta'
e^{-\gamma_0\delta'}
\left[
1-\exp\left(
  -p_1\,e^{\gamma_0\delta'}
  \right)
\right]
\int_0^{\tilde y_0}d\tilde y_1
\left(
  \frac{y}{y_1\tilde y_1}
  \right)^{3/2}
\exp\left(
-\frac{{\delta'}^2}{2\chi''(\gamma_0)\tilde y_1}
\right).
\ee
The $\tilde y_1$ integration can be performed up to a correction
of order $1/\sqrt{y}$, noticing that the integral is dominated
by the region $\tilde y_1\ll y$, hence $y_1\simeq y$.
One can in particular replace the upper
bound by~$\infty$:
\be
\int_0^{\infty}\frac{d\tilde y_1}{\tilde y_1^{3/2}}
\exp\left(
-\frac{{\delta'}^2}{2\chi''(\gamma_0)\tilde y_1}
\right)=\frac{\sqrt{2\pi\chi''(\gamma_0)}}{\delta'}.
\ee
What remains is an integration over $\delta'$:
\be
T_1(y,x)={C}\sqrt{2\pi\chi''(\gamma_0)}
\int_0^\infty d\delta'\,
e^{-\gamma_0\delta'}
\left[
1-\exp\left(
  -p_1\,e^{\gamma_0\delta'}
  \right)
  \right].
\label{eq:T1front}
\ee
This expression can be interpreted as the scattering
amplitude of an onium the state of which
was determined by a {\it front fluctuation} in
the terminology of Ref.~\cite{Mueller:2014gpa,Mueller:2014fba},
which is meant to be a fluctuation in the beginning of the evolution
that essentially shifts the whole dipole distribution towards larger
sizes. The size of this class of fluctuations was assigned an exponential distribution
$\propto e^{-\gamma_0\delta'}$ for large $\delta'$:
This is precisely the weight of $\delta'$
that appears in the remaining integration.

Equation~(\ref{eq:T1front}) may be rewritten with the help of
the integral $I_1$ defined and evaluated in Appendix~\ref{app:integral}:
\be
T_1(y,x)=
\frac{C}{\gamma_0}\sqrt{2\pi\chi''(\gamma_0)}\times I_1(p_1).
\ee
Using Eq.~(\ref{eq:I1}) and replacing $p_1$ by its
definition in Eq.~(\ref{eq:I}),
the final result reads
\be
T_1(y,x)\simeq CC_1C_2\,\pi\left[\chi''(\gamma_0)\right]^2
(x-X_y)\,e^{-\gamma_0(x-X_y)}.
\label{eq:T1final}
\ee
We have just recovered the scaling limit of the known solution to the
BK equation, see Eq.~(\ref{eq:T1BK}).
We have assumed throughout $(x-X_y)^2\ll y$,
so we cannot get consistently the finite-$y$
corrections that appear in the form of an exponential of the ratio
of the first over the second scale multiplied by
a negative constant factor.
We have learned from this calculation that, in the considered frame,
the realizations of the Fock states which trigger events look like typical
realizations, as far as their shapes is concerned, but overall shifted towards larger
dipole sizes by a multiplicative factor, through a fluctuation occurring
at the very beginning of the evolution.


\subsubsection{Multiple scatterings: $G$ and $T_2$}

As for the calculation of $G$, we start with Eq.~(\ref{eq:Gpheno}),
and substitute $I$ by the expression obtained in Eq.~(\ref{eq:I}):
\begin{multline}
G(y,x;y_1)=C
\int_0^{\delta_0} d\delta\,\delta\,e^{-\gamma_0\delta}
\exp\left(
-\frac{\delta^2}{2\chi''(\gamma_0)\tilde y_1}
\right)\\
\times
\left\{
  1-\left[1+p_1 \left(
\frac{y}{y_1\tilde y_1}
\right)^{3/2}
e^{\gamma_0\delta}\right]
  \exp\left[-p_1 \left(
\frac{y}{y_1\tilde y_1}
\right)^{3/2}
e^{\gamma_0\delta}\right]
  \right\}.
  \label{eq:Gstep}
\end{multline}
Again, because of the form of the integrand,
$\delta$ does not exceed $x-X_y$. This implies that the upper bound $\delta_0$
can be replaced by $+\infty$.
Furthermore, $x-X_y$ is assumed, as a consequence of our choice of frame, to
be much less than $\sqrt{y_0}$. Hence if one restricts oneself to
values of $\tilde y_1\equiv y-y_1$ not smaller than $y_0$, the Gaussian
factor can be set to~1.
Then, performing the change of variable $t\equiv e^{\gamma_0\delta}$,
$G$ boils down to an integral computed in Appendix~\ref{app:integral}, up to an
overall factor:
\be
G(y,x;y_1)=\frac{C}{\gamma_0^2}
\times I'_2
\left[
p_1\left(\frac{y}{y_1\tilde y_1}
\right)^{3/2}
\right].
\ee
It follows that
\be
\begin{split}
G(y,x;y_1)&\simeq\frac{C}{\gamma_0^2}\times p_1\left(\frac{y}{y_1\tilde y_1}\right)^{3/2}
\ln \left[\frac{1}{p_1}
  \left(\frac{y_1\tilde y_1}{y}
\right)^{3/2}
  \right]\\
&\simeq\frac{CC_1C_2}{\gamma_0}\sqrt{\frac{\pi}{2}}
\left[\chi''(\gamma_0)\right]^{3/2}
\left(\frac{y}{y_1\tilde y_1}\right)^{3/2}
     {(x-X_y)\,e^{-\gamma_0(x-X_y)}},
\end{split}
\label{eq:Gfinal}
\ee
where we neglected additive constants and slowly varying logarithms of the
rapidities, which are small compared to $x-X_y$.
Since $(x-X_y)\,e^{-\gamma_0(x-X_y)}\simeq{T_1(y,x)\times
  \left\{CC_1C_2\pi[\chi''(\gamma_0)]^2\right\}^{-1}}$
(see Eq.~(\ref{eq:T1final})),
we arrive at the following expression for the distribution
of $y_1$ normalized to the amplitude $T_1$:
\be
{
\frac{G(y,x;y_1)}{T_1(y,x)}
=\frac{1}{\gamma_0}\frac{1}{\sqrt{2\pi\chi''(\gamma_0)}}
\left(\frac{y}{y_1(y-y_1)}\right)^{3/2}.
}
\label{eq:GoverT1}
\ee

We see that the small-$\tilde y_1$ region is highly singular, but
the singularity has to be cut off by a factor that is subleading when
$\tilde y_1$ is taken on the order of $y$. This region would correspond to
the production of a large dipole in the very beginning of the evolution
of the onium, but this is suppressed: Indeed, 
the mechanism leading to a particle away from the mean
position of the lead particle is diffusive, and the diffusion radius grows like
$\sqrt{\tilde y_1}$.
Consequently, one may expect the expression~(\ref{eq:GoverT1})
to be supplemented by a multiplicative factor ${\cal D}(x-X_y,y-y_1)$,
where
\be
   {\cal D}(\Delta X,\Delta y)\equiv
   \exp\left(-\frac{\Delta X^2}{2\chi''(\gamma_0)\Delta y}\right).
\label{eq:diffusive}
\ee

The scattering amplitude conditioned to having at least two scatterings between
the state of the onium evolved to rapidity $\tilde y_0$ and the nucleus,
$T_2$, is an integral of $G$ over $y_1$.
Starting with Eq.~(\ref{eq:Gstep}), its evaluation goes along the same lines
as that of $T_1$ above.
The $y_1$-integration can be performed in the first place.
We are then left with an integral over $\delta$, which takes the
form
\be
T_2(y,x;y_0)=\frac{C}{\gamma_0}\sqrt{2\pi\chi''(\gamma_0)}\times I_2(p_1).
\ee
Using the evaluation of $I_2$ in Appendix~\ref{app:integral}, Eq.~(\ref{eq:I2}),
and replacing $p_1$ by its expression~(\ref{eq:def-p_1}), we find a very
simple relation between $T_1$ and $T_2$ in this frame:
\be
{
  \frac{T_2(y,x;y_0)}{T_1(y,x)}
  \underset{y_0\gg (x-X_y)^2}{=}
  \frac{1}{\gamma_0(x-X_y)}.
  }
\ee
It turns out that we would have got the same result by
integrating Eq.~(\ref{eq:GoverT1}) supplemented with the diffusive
factor ${\cal D}(x-X_y,\tilde y_1)$ defined in Eq.~(\ref{eq:diffusive})
which cuts off the very small $y_1$ region.

We note that the overall constant in front of the ratios
$T_2/T_1$ and $G/T_1$ is sensitive to the detailed form of the
interaction between the dipoles and the nucleus, which was not the case
for $T_1$. Within the assumptions of the phenomenological model,
this form is unambiguous:
The number of scatterings at the time of the interaction
obeys a Poisson law of parameter~$I$.
Whether the overall constant found in this model
is the correct one for
branching random walks and for the QCD dipole model
depends on the ability of the phenomenological model to capture
accurately enough the features of the latter models:
We will need numerical calculations to check it (see Sec.~\ref{sec:numerics} below).

Finally, for the same arguments as the ones presented in Sec.~\ref{sec:model-dipole-dist},
the assumption that no single dipole has a significant probability to scatter
also proves correct a posteriori.


\subsection{What happens in a frame in which the nucleus is less boosted?}

The choice of the reference frame was very important in the calculation
above: We chose a frame in which the nucleus is highly boosted, such that
$y_0\gg (x-X_y)^2$. Such a choice implies that the scattering
configurations are dominated by fluctuations which occur very
early in the rapidity evolution.
This is a perfectly valid choice, as long as
we pick $x$ in the scaling region, i.e. $(x-X_y)^2\ll y$.
However, any other frame should be allowed.
We shall investigate the scattering picture in frames in which the
nucleus is at rest or close to rest.

\subsubsection{Nucleus restframe}
The amplitude $T_1$ was analyzed in the
restframe of the nucleus ($y_0=0$)
in Ref.~\cite{Mueller:2014fba}.
In that frame, the nucleus has not developed a universal front: The scattering
amplitude of a dipole becomes very small as soon as the size of this dipole
gets smaller than the inverse saturation scale $1/Q_A$; see Eq.~(\ref{eq:MV}).
Therefore, in
all events, the fluctuations of the partonic content of the onium
must produce at least one dipole which will
be completely absorbed by the nucleus, namely, which has a size larger
than $1/Q_A$. The formulation of $T_1$ simply reads
\be
\left.T_1(y,x)\right|_{y_0=0}=\int dx' p(x-x',y) T_1(0,x'),
\ee
where $T_1(0,x')=1-S(0,x')$
is the McLerran-Venugopalan amplitude given in Eq.~(\ref{eq:MV})
that we may approximate by a Heaviside distribution with
support the set of negative real numbers.

The leading term in the integral over $x'$ can then be obtained
quite easily. The integration over $x'$ is dominated
by a region of log-size of order 1 around $x'=0$.
The result is proportional to
\be
\left.T_1(y,x)\right|_{y_0=0}\propto
c(x-X_y)e^{-\gamma_0(x-X_y)},
\ee
which is of course what is expected at the parametric level.
However,
the overall constant $c$ cannot be easily related to $C$, $C_1$
and $C_2$ of the phenomenological model.
This is because the latter are unambiguously defined for
an evolved front, once a convention for the
definition of the front position/saturation scale has been
chosen, but the transition between the initial
condition and the well-developed front is not controled analytically
in the initial stages of the evolution.

$T_2$ and $G$ cannot be calculated in this frame.
Indeed, their evaluation requires the precise understanding of
the particle distribution in fluctuations happening near the boundary
of the BRW, which is still an unsolved problem.


\subsubsection{Slightly boosted nucleus}

We now investigate the case of the frame in which the nucleus is
boosted only slightly. We shall choose a frame in which the
rapidity of the nucleus satisfies, parametrically,
\be
1\ll y_0\ll (x-X_y)^2.
\ee
While for the opposite ordering between $y_0$
and $(x-X_y)^2$ one could perform
a relatively straightforward calculation, this case is much trickier.
We shall show how the calculations of $T_1$ and $T_2$
go, which will enable us to understand what the
typical state of the onium looks like when viewed from
this particular frame.

\paragraph{Calculation of $T_1$.}
Anticipating that the main contribution will come from the
configurations which do not overlap with the
saturation region of the nucleus, we expand the exponential
in Eq.~(\ref{eq:T1pheno}):
\be
\left.T_1(y,x)\right.
=\int_{y_0}^{y} dy_1\int_0^{\delta_0}d\delta\,
p(\delta,\tilde y_1)\,I(y_0;\delta,y_1)\Theta(x-X_y-\delta).
\label{eq:T1phenotricky}
\ee
Substituting $I$, we get
\begin{multline}
  \left.T_1(y,x)\right.
  =\frac{CC_1C_2}{y_0^{3/2}} e^{-\gamma_0(x-X_y)}
  \int_0^{x-X_y} d\delta\,\delta
    \int_{y_0}^{y} dy_1
\int_{\tilde X_{\tilde y_0-\tilde y_1}+\Xi_{\delta,\tilde y_1}}^{\infty} dx'
\,(x'-X_{y_0})\,
\underbrace{\exp\left(
-\frac{(x'-X_{y_0})^2}{2\chi''(\gamma_0) y_0}
\right)}_{\text{(I)}}\\
\times
\left(
  \frac{y}{\tilde y_1(\tilde y_0-\tilde y_1)}\right)^{3/2}
  (x'-\tilde X_{\tilde y_0-\tilde y_1}-\Xi_{\delta,\tilde y_1})\,
\underbrace{\exp\left(
-\frac{(x'-\tilde X_{\tilde y_0-\tilde y_1}-\Xi_{\delta,\tilde y_1})^2}
{2\chi''(\gamma_0) (\tilde y_0-\tilde y_1)}
\right)}_{\text{(II)}}
\underbrace{\exp\left(
-\frac{\delta^2}{2\chi''(\gamma_0) \tilde y_1}
\right)}_{\text{(III)}}.
\end{multline}
It is not possible to perform these nested integrals
exactly, but we can extract the asymptotic expression of
$T_1$ in a definite limit.

The Gaussian factors~(I),(II),(III) set effective cutoffs: Their analysis
enables us to assess which subdomains of the integration
region will give the dominant contribution, and thus to judge which
approximations we may afford without altering the value of the integral
in the asymptotic limits of interest here.

The presence of the factor~(I)
implies that $x'-X_{y_0}$ must be at most of order $\sqrt{y_0}$. Since
$x'$ is larger than the position of the leftmost tip of the
dipole distribution,
$\Delta(y_0;\delta,y_1)$ defined in Eq.~(\ref{eq:Delta_def})
is also at most of order $\sqrt{y_0}$. This in turn implies that
the size of the fluctuation be $\delta\sim x-X_y$,
up to ${\cal O}(\sqrt{y_0})$.
The factor (II) forces $\tilde y_0-\tilde y_1\lesssim y_0$, i.e.
$\tilde y_1\simeq y$ up to corrections of order $y_0$.
So whenever a factor $\tilde y_1$ appears one can safely replace it by $y$.
The last factor (III) is always of order unity since
$\delta$ is of order $x-X_y$, and since
we have chosen to stick to the scaling
region, defined by $x-X_y \ll \sqrt{y}\sim\sqrt{\tilde y_1}$.
We further observe that the $\tilde y_1$-dependence of
$\tilde X_{\tilde y_0-\tilde y_1}+\Xi_{\delta,\tilde y_1}$
is only logarithmic, hence it can be neglected here, given that
these terms do not appear in exponential factors which would have
enhanced their contribution.
In other words, one can afford the approximation
\be
\Delta(y_0;\delta,y_1)\simeq x-X_{y}-\delta.
\ee
Once this approximation is implemented, one
can perform the integral over $y_1$.
Taking into account that the dominant region is such that $y_1$
is close to $y_0$, we set the upper boundary to $+\infty$,
and replace $y/\tilde y_1$ by 1. 
It just gives a number:
\be
\int_{y_0}^{+\infty}\frac{dy_1}{(\tilde y_0-\tilde y_1)^{3/2}}
(x'-\tilde X_{\tilde y_0-\tilde y_1}-\Xi_{\delta,\tilde y_1})
  \exp\left(
-\frac{(x'-\tilde X_{\tilde y_0-\tilde y_1}-\Xi_{\delta,\tilde y_1})^2}
{2\chi''(\gamma_0) (\tilde y_0-\tilde y_1)}
\right)\simeq \sqrt{2\pi\chi''(\gamma_0)}.
\ee
Further, the integral over $x'$ boils down to the integral of an
exponential. After the change of integration variable $\bar x\equiv x-X_{y_0}$,
it reads
\be
\int_{x-X_{y}-\delta}^{+\infty}d\bar x\, \bar x
\exp\left(
-\frac{\bar x^2}{2\chi''(\gamma_0) y_0}
\right)=\chi''(\gamma_0)y_0\times
\exp\left(-\frac{(x-X_y-\delta)^2}{2\chi''(\gamma_0)y_0}
\right).
\ee
Finally, the integration over $\delta$ is dominated by a
region of size $\sqrt{y_0}$ close to $x-X_y$, namely
\be
\int_0^{x-X_y} d\delta\,\delta\,
\exp\left(-\frac{(x-X_y-\delta)^2}{2\chi''(\gamma_0)y_0}\right)
\simeq \sqrt{\frac{\pi}{2}}\sqrt{\chi''(\gamma_0)y_0}
\times(x-X_y).
\ee
Putting all factors together, we get Eq.~(\ref{eq:T1final}),
the overall multiplicative constants being identical:
\be
\left.T_1(y,x)\right|_{y_0\gg (x-X_y)^2}=
\left.T_1(y,x)\right|_{1\ll y_0\ll (x-X_y)^2}.
\ee
Hence we have checked explicitly that boost invariance holds.

Interestingly enough, the physical pictures in both frames are very different.
Indeed, from the
analysis of the integration domain, we see that in the frame
in which ${y_0}\ll (x-X_y)^2$, the fluctuation
occurs typically late in the onium evolution,
at rapidity $\tilde y_1$ close to the scattering rapidity $\tilde y_0$.
It takes place in a window
of size of order $y_0$ in such a way that the overlap with the front
of the nucleus, of size $\sqrt{y_0}$, may be significant.
This is necessary since the fluctuation
needs to extend far out of the ``mean field'' region,
and thus requires a large rapidity range to develop.
The particle front that interacts with the nucleus,
which results from the evolution of the fluctuation, is of
size $\sqrt{y_0}$,
that is, it has just the right size to have an optimal overlap with
the front of the nucleus.

\paragraph{Calculation of $T_2$.}

This quantity is significantly more difficult to compute.
First, unlike in the case of $T_1$, expanding the exponential
in Eq.~(\ref{eq:T2pheno}) is not licit, and leads to a
loss of control of the constant factors multiplying the
leading term.

What happens physically is that requiring at least two scatterings
pushes $\delta$ to take a value for which $I(\delta,y_1)\sim 1$ in each
event, which limits the possible values of $\delta$ to a narrow interval.
Hence the integral over $\delta$ does not bring a factor $\sqrt{y_0}$
reflecting the size of the integration region.
This is essentially the difference between the calculation of $T_1$
and that of $T_2$ in this regime. This reasoning leads to the following
estimate of its parametric form:
\be
\left.T_2(y,x;y_0)\right|_{1\ll y_0\ll (x-X_y)^2}\sim\frac{T_1(y,x)}{\sqrt{y_0}}.
\ee

In order to get a more complete expression, we
may recognize that $G/T_1$ is boost-invariant, and integrate
its expression obtained in Eq.~(\ref{eq:GoverT1}),
supplemented with the Gaussian factor~(\ref{eq:diffusive}), over $y_1$.
In this case, the integral is dominated by the
region close to $y_1\sim y_0$, i.e. $\tilde y_1\simeq y$.
The result has the same parametric form as the one just guessed,
but the overall constant is completely determined:
\be
{
\frac{T_2(y,x;y_0)}{T_1(y,x)}\underset{1\ll y_0\ll (x-X_y)^2}{=}
\frac{1}{\gamma_0}\sqrt{\frac{2}{\pi\chi''(\gamma_0)}}
\frac{1}{\sqrt{y_0}}.
}
\ee


\section{\label{sec:numerics}
  Comparing the model predictions with the solutions
  to the exact equations}

We shall now check the results we have obtained using the phenomenological
model by solving numerically the exact equations.
In order to compare more easily different values of $y$, it is useful
to introduce the overlap $q\equiv \tilde y_1/y$, representing the
fraction of the total rapidity over which there is a unique common ancestor
of the dipoles that eventually interact.
Its distribution is just given by $G/T_1$, up to the change of
variable and the corresponding Jacobian:
\be
{\pi}_\infty(q)=\frac{1}{\sqrt{y}}\frac{1}{\gamma_0\sqrt{2\pi\chi''(\gamma_0)}}
\frac{1}{q^{3/2}(1-q)^{3/2}}.
\label{eq:overlap}
\ee
The $y$-dependence of this asymptotic result is very simple,
consisting only in the multiplicative factor $1/\sqrt{y}$: Therefore,
we shall keep it implicit in the definition of the asymptotic distribution.
The ``$\infty$''
subscript reminds that this expression is valid for asymptotic values of $y$.

We shall not use the QCD dipole model, but the simpler branching random walk
introduced in Ref.~\cite{Anh:2020bss} and further investigated
in Ref.~\cite{Brunet_2020}.
Indeed, it is know that the form of the asymptotics is the same for
all models in the universality class of branching diffusion.
Only the parameters $\gamma_0$, $\chi'(\gamma_0)$, and $\chi''(\gamma_0)$,
which depend on the detailed elementary processes,
need to be substituted.


\subsection{Definition of the implemented model}

We consider a branching random walk in discrete space and time, defined
by the following processes: Between the rapidities $y$ and $y+\delta y$,
a particle on site $x$ may jump to the site on the
left (i.e. at position $x-\delta x$) or on the right ($x+\delta x$)
with respective probabilities $\frac12(1-\delta y)$, or
may branch into two particles on the same site $x$
with probability $\delta y$.

The main differences with respect to the QCD dipole model
is that in the latter,
the diffusion and the branching actually 
happen at the same time through a single process,
and that QCD is a theory in the continuum.
But these differences should not affect
the asymptotics of the observables we are considering.

The fundamental quantity for us is the probability
that there is no particle to the right of the site at some position $X$. This 
is the equivalent of the $S$-matrix element in the QCD case.
It evolves in rapidity according to the equivalent of
the BK equation~(\ref{eq:BK}) for this model,
which is the finite-difference equation
\be
S(y+\delta y,x)=\frac12(1-\delta y)\left[ S(y,x-\delta x)+S(y,x+\delta x)\right]
+\delta y\left[S(y,x)\right]^2,
\label{eq:equivBK}
\ee
with the initial condition $S(y=0,x\leq 0)=0$ and $S(y=0,x>0)=1$.
In the numerical calculation, we shall take the following values for the
parameters:
\be
\delta y=0.01,\quad\delta x=0.1.
\ee
The values of $\gamma_0$, $\chi'(\gamma_0)$, $\chi''(\gamma_0)$ are
obtained from the general solution to Eq.~(\ref{eq:equivBK})
linearized near $S\sim 1$. For this model, we find~\cite{Brunet_2020}
\be
\gamma_0=1.4319525\cdots,\quad
\chi'(\gamma_0)=1.3943622\cdots,\quad
\chi''(\gamma_0)=0.96095291\cdots.
\label{eq:par1}
\ee


\subsection{Numerical calculation of the distribution of the splitting
rapidity of the parent dipole}

\begin{figure}
  \begin{center}
    \includegraphics[width=.9\textwidth]{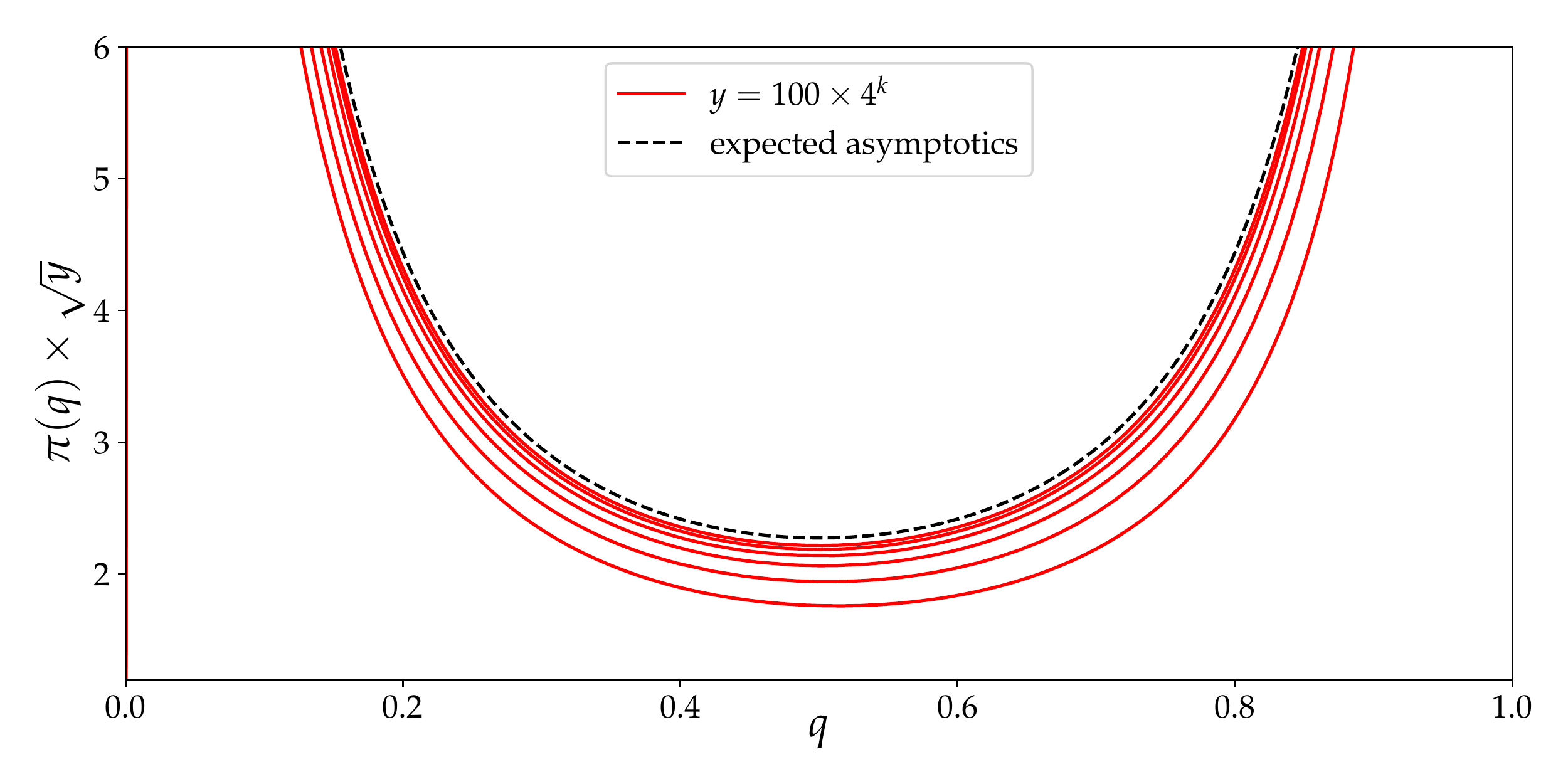}
  \end{center}
  \caption{{\small Distribution of the overlaps for different values of $y$
    (set of full lines; $y=100\times 4^k$ with $k=1,\cdots,6$ increasing
    from bottom to top) and $X-X_y$ set to $\sqrt{2}\,y^{1/4}$,
    together with the expected asymptotics given in Eq.~(\ref{eq:overlap})
    (dashed line).}
    \label{fig:ancestry_11}}
\end{figure}

The equation to solve to get the equivalent of $G$ in the framework of this
model is the following:
\be
G(y+\delta y,x;y_1)=\frac12(1-\delta y)\left[G(y,x-\delta x;y_1)+G(y,x+\delta x;y_1)\right]
+2\delta y \,G(y,x;y_1)\,S(y,x),
\label{eq:evolGmodel}
\ee
with
\be
G(y_1,x;y_1)=\left[1-S(y_1,x)\right]^2.
\label{eq:initGmodel}
\ee
(Compare to Eq.~(\ref{eq:evolG}) and~(\ref{eq:initG}) respectively
obtained in the dipole model).

In order to satisfy in some optimal way the double constraint
in Eq.~(\ref{eq:definition-scaling-region}),
we set $x$ to a value $X$ such that 
\be
X\simeq X_y+\sqrt{\kappa}\, y^{1/4},
\ee
where $\kappa$ is a constant that we shall pick in the set $\{1,2,4\}$.
In general, we cannot achieve the equality because $X$ is the position of a lattice
site, so it is discrete, while the r.h.s. is a real number: We have picked the closest 
site to the left of the position in the r.h.s.
Up to a numerical factor, the r.h.s. is the geometric average of the two
bounds on $X-X_y$ in Eq.~(\ref{eq:definition-scaling-region}). Varying
this constant enables one to go more or less deep in the scaling region.

We have collected data for $y$ up to ${\cal O}(10^6)$.
We plot the distribution
$\pi_y(q)$ at finite rapidity rescaled by $\sqrt{y}$ for different $y$ and $\kappa$
in Fig.~\ref{fig:ancestry_11}, together with the expected infinite-$y$ asymptotic
distribution $\pi_\infty(q)$.

\begin{figure}
  \begin{center}
    \includegraphics[width=.9\textwidth]{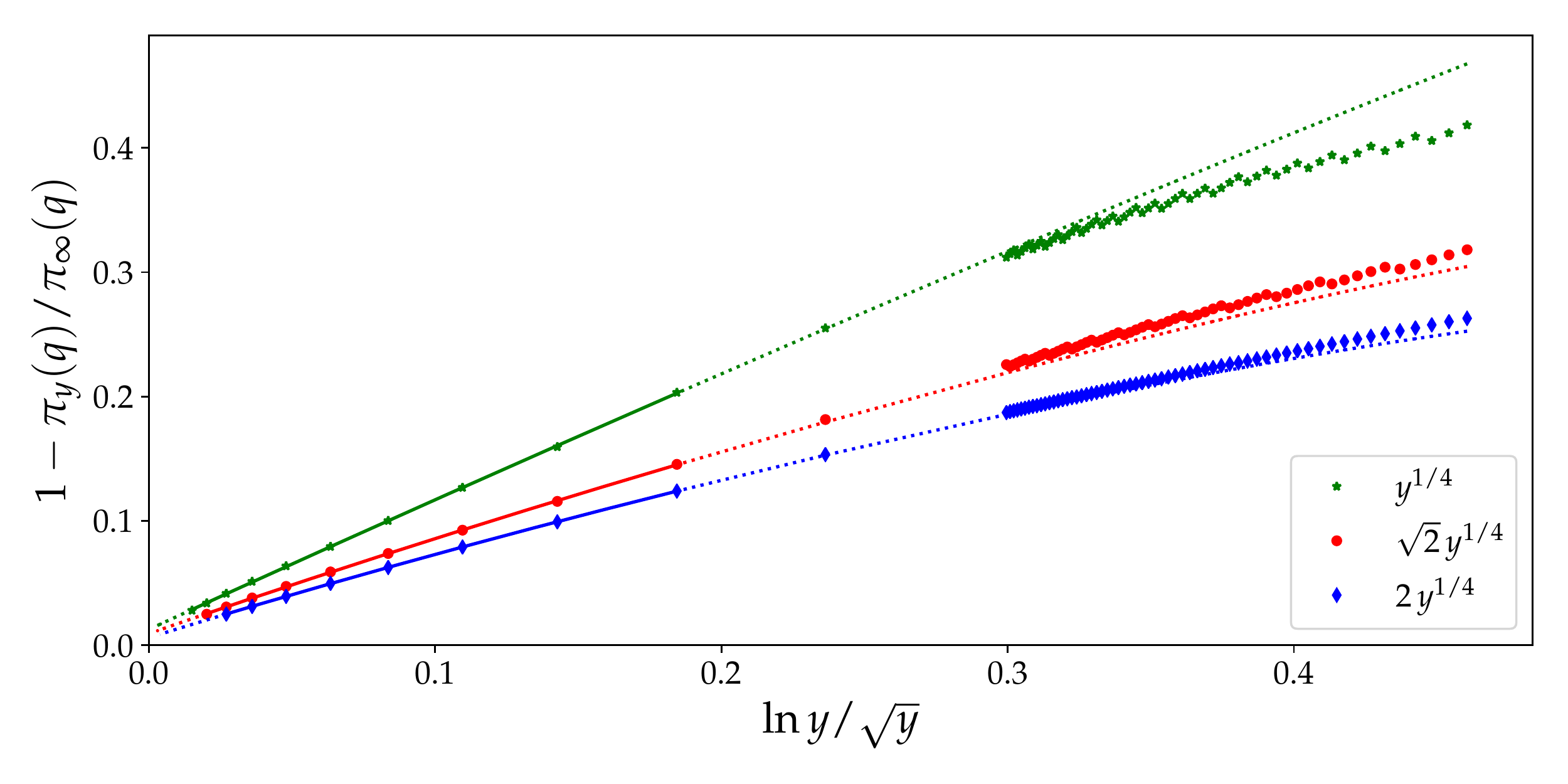}
  \end{center}
  \captionof{figure}{{\small Complementary to 1 of the ratio between
      the probability of an overlap $q=0.5$ for different values
      of the total rapidity~$y$ as a function of $\ln y/\sqrt{y}$.
    The points stem from the numerical integration of the evolution
    equations~(\ref{eq:equivBK}),(\ref{eq:evolGmodel}),(\ref{eq:initGmodel}).
    The continuous lines, meant to guide the eye,
    represent the function~(\ref{eq:fit}) fitted to these points, and
    the dotted lines are extrapolations outside the domain in which
    the fit is performed.}
    \label{fig:ancestry_convergence}}
\end{figure}

In order to appreciate the convergence
quantitatively, we pick a point of fixed $q$,
and we compare the measured $\pi_y(q)$ at finite
rapidity to the expected one $\pi_\infty(q)$ at $y=\infty$.
In practice, we have chosen $q=0.5$, but we have also tried other values and
got similar results for this ratio.
We plot the complementary to one of this
ratio against $\ln y/\sqrt{y}$ in Fig.~\ref{fig:ancestry_convergence}.
We expect a curve that
goes through the origin: We see
a quite good convergence of $1-\pi_y(q)/\pi_\infty(q)$
to 0 when $y\rightarrow\infty$.

Note that some non-smooth structures appear. They just reflect
the discreteness of the model, which forces us to set $X$ to discrete values
according to the procedure described above.
The discretization step in $x$ is
$\delta x=0.1$ in this model, which is not that small compared to
$y^{1/4}$, even for very large $y$. So moving $X$ by $\delta x$
can lead to sizable differences in $\pi_y(q)$. Of course,
these differences should get smaller and smaller
as $y\rightarrow\infty$.

In order to guide the eye, we superimpose
the following function to the data:
\be
1-\frac{\pi_y(q=0.5)}{\pi_\infty(q=0.5)}
\underset{\text{fit}}{=}
a+\frac{b_{11}\ln y+b_{12}}{\sqrt{y}}.
\label{eq:fit}
\ee
We fit the parameters $a$, $b_{11}$, $b_{12}$,
to the numerical calculations.
We get reasonable values for all these parameters for
the considered choices of $X$:
$b_{11}$ and $b_{12}$ are of order~1, while $a$ is found close to zero;
see Tab.~\ref{tab:fit}.
The constant $a$ is of the order of a percent, when we expect it to vanish.
But we can hardly aim for better, because of the
structures induced by the discreteness of the model, which makes the fitting
procedure by a smooth function dependent on the choice of the points.
Therefore, we conclude that our analytical formula~(\ref{eq:overlap}) is
well-supported by this numerical calculation.

However, these results show that the finite-$y$ corrections are definitely
very large, and the convergence slow: Although we have computed
$\pi_y$ for values of $y$ on the order of
$10^6$, we have not managed to approach the asymptotics by better than about $3\%$.
Figure~\ref{fig:ancestry_convergence} and the fitted formula~(\ref{eq:fit})
seem to indicate that the
correction to $\pi_\infty(q)$ may take the form of a multiplicative factor
$\left(1+\text{const}\times\ln y/\sqrt{y}\right)$. But we have no
theory that may enable us neither to understand nor to guess
the form of $\pi_y(q)$  beyond the leading term
in the limit of large $y$.

We have also implemented independently
another branching random walk model, and we have reached the same conclusion; see
Appendix~\ref{app:num} for a presentation of the model and of the obtained results.

\begin{table}
  \begin{center}
  \begin{tabular}{c||c|c|cccc}
    \hline
    $X-X_y$ & $a$ ($\times 10^{-2}$) &
    $b_{11}$ & $b_{12}$ \\
    \hline
    $y^{1/4}$           & 1.26 & 1.10 & -0.519 \\
    $\sqrt{2}\,y^{1/4}$ & 0.909 & 0.896 & -1.17 \\
    ${2}\,y^{1/4}$      & 0.587 & 0.813 & -1.28 \\
    \hline
  \end{tabular}
  \captionof{table}{\small
    Values of the parameters in Eq.~(\ref{eq:fit}) obtained from
    a fit to the numerical data shown in Fig.~\ref{fig:ancestry_convergence}.
    \label{tab:fit}}
  \end{center}
\end{table}


\section{\label{sec:conclusion}Summary and outlook}

In onium-nucleus scattering, in a frame in which the onium moves with a large
rapidity $\tilde y_0$, the latter interacts through a typically
dense quantum state made of gluons, that we represented by a set of color dipoles.
Only a subset of these dipoles actually exchange energy with the nucleus. Having
at least one dipole in this set is necessary to have a scattering event:
This requirement defines the forward elastic scattering amplitude $T_1$.
Calculating the joint probability $T_2$ to have a scattering and at least two dipoles in the
set makes possible to get quantitative information on the correlations of
the dipoles involved in the interaction.

In this paper, we have shown that the shape of the partonic configurations of an
onium that interacts with a large nucleus
depends qualitatively on the chosen reference frame.
If the nucleus is highly boosted, namely if its rapidity
$y_0$ is much larger than $\ln^2 [r^2Q_s^2(y)]$, then the dipole distribution
at the time of the interaction with the nucleus looks like a typical (``mean-field'')
distribution just shifted (through a ``front fluctuation'' occurring in
the very beginning of the evolution) towards larger sizes.
If instead the nucleus is less boosted, $1\ll y_0\ll\ln^2 [r^2Q_s^2(y)]$,
then the dipoles which interact with it stem from a tip fluctuation
occurring at much larger rapidities $\tilde y_1$ in the onium evolution,
of order $y$ such that $y_1\sim y_0$.

\begin{figure}
  \begin{center}
    \includegraphics[width=.9\textwidth]{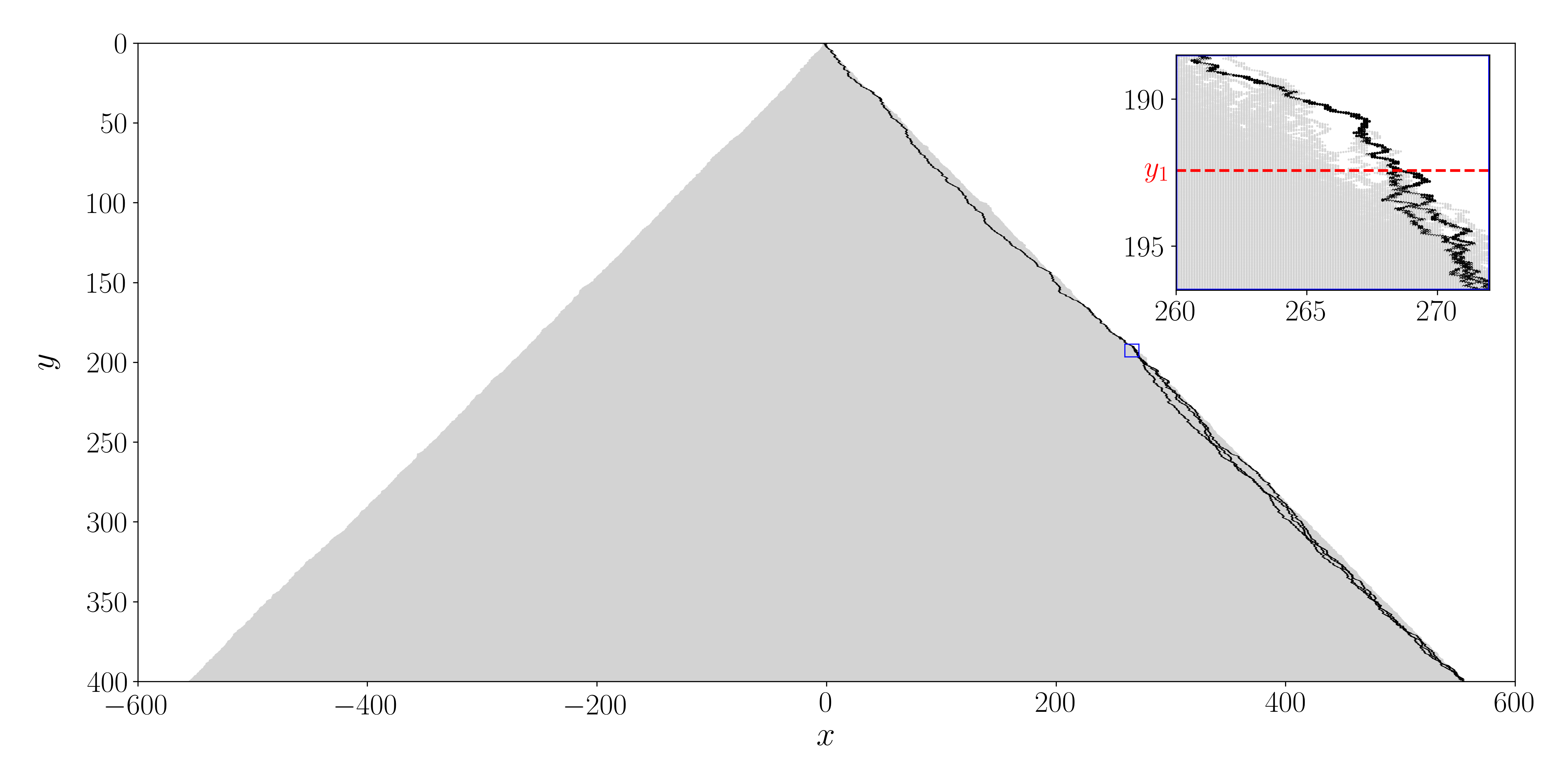}
  \end{center}
  \caption{\label{fig:event}{\small
      One particular realization of the toy model described in Sec.~\ref{sec:numerics}
      evolved up to $y=400$, that turns out
      to possess several particles to the right of the position
      $X\simeq X_{y=400}+\sqrt{2}\times{400}^{1/4}$.
      The grey zone is the set of non-empty lattice sites for all values of the rapidity.
      The black lines
      represent the worldlines of all the particles that end up with a position
      not less than $X$ at the final rapidity $y=400$.
      The common ancestor of these particles splits at $y_1=192.43$.
      The inset is a zoom on the branching
      region around the branching rapidity $y_1$,
      illustrating that this common ancestor indeed stems from a large fluctuation
      occurring at a rapidity close to $y_1$, as assumed in the phenomenological model.
      This rare realization was generated using the algorithm proposed in Ref.~\cite{Brunet_2020}.
    }
    }
\end{figure}

Choosing a frame such that the tip fluctuation, the offspring of which
scatter with the nucleus, be sufficiently developed,
we were able to calculate the asymptotics
of the distribution of the splitting rapidity of the slowest ancestor
of the set of dipoles that effectively interact with the nucleus,
including the overall constant.
We have found that its ratio 
to the total rapidity of the scattering, a quantity that we denote by $q$,
is distributed as
\be
{\pi}_\infty(q)=\frac{1}{\sqrt{y}}\frac{1}{\gamma_0\sqrt{2\pi\chi''(\gamma_0)}}
\frac{1}{q^{3/2}(1-q)^{3/2}}.
\label{eq:overlap_f}
\ee
This expression holds when the size $r$ of the onium is chosen in the so-called
scaling region, defined as
\be
1\ll \ln^2 \frac{1}{r^2Q_s^2(y)}\ll y,
\quad\text{namely}\quad
1 \ll (x-X_y)^2 \ll y,
\label{eq:definition-scaling-region-f}
\ee
where $x$ and $X_y$ just correspond to $1/r$ and $Q_s(y)$
respectively when measured on a logarithmic scale (see Eq.~(\ref{eq:notation_log-size})),
and $X_y=\chi'(\gamma_0) y-\frac{3}{2\gamma_0}\ln y$.
Equation~(\ref{eq:overlap_f})
is our main quantitative result.
A particular realization of the model introduced in Sec.~\ref{sec:numerics}
and used to check numerically the
calculations is shown in Fig.~\ref{fig:event}.

After identification of the rapidity with a time variable,
we expect this expression to represent the distribution
of the relative branching time of the most recent common ancestor of all particles
that end up to the right of some predefined position $x$
for any branching random walk -- provided that $x$ is chosen
in the scaling region. (This $q$ is also called ``overlap''
in the statistical physics literature). The constants that appear
in these expressions are easily calculated from the
detailed form of the elementary processes which define the branching random
walk.

We observe that our result coincides with a conjecture by Derrida
and Mottishaw~\cite{DerridaMottishaw} for a slightly different 
genealogy problem in the context of general branching
random walks: They computed the distribution of the
branching time of the most recent common ancestor of two particles
of predefined order number counted from the tip of the particle distribution
at some given large time.
To make contact between our equation~(\ref{eq:overlap_f}) and the formula~(6)
they wrote down in Ref.~\cite{DerridaMottishaw},
we just need to identify our $\pi(q)$ with their $p(q)$, $y$ with the total evolution
time $t$, $\gamma_0$ with $\beta_c$, and $\chi''(\gamma_0)$ with
$\beta_c\, v''(\beta_c)$. The constant $\chi'(\gamma_0)$, that enters the validity
condition, is just to be identified with
the critical FKPP front velocity $v(\beta_c)$.

While this expression was established in \cite{DerridaMottishaw}
through a calculation in
the framework of the Generalized Random Energy Model \cite{GREM},
we have derived it in the context of the problem we were addressing
from the phenomenological model for
branching random walks, which just assumes that the time evolution
is essentially deterministic, except for one single fluctuation.
Proving our result for the genealogies rigorously,
establishing a framework for the systematic calculation of corrections,
crucial for applications since the approach to the asymptotics turns out
to be very slow, are problems of general interest for branching processes,
and exciting challenges for further investigations.

Also, our
method would not apply to the specific genealogy problems Derrida-Mottishaw were
addressing: While the $y$ (or $t$)-dependence would turn out correct,
the overall constant could not be obtained.
The reason for this is that we do not have a sufficient understanding of the particle
distributions and of their correlations very close to the lead particle.
Trying to build a good picture of the latter~\cite{Brunet2011,Anh:2020bss,Brunet_2020}
is a long-term program that deserves more efforts.

As for the more specialized diffraction problem, which was the initial motivation
for the present work,
we also intend to try and extend our calculation to the rate of rapidity
gaps in high-energy onium-nucleus scattering. The main crucial difference is that the
latter being a quantum mechanical
observable, it has no interpretation in purely classical probabilistic terms.
However, preliminary investigations
seem to indicate that the technical tools developed here can be applied also
to that observable, which will be measurable at a future electron-ion collider.
Of course, the finite-rapidity $y$ subasymptotics (presumably of relative order $\ln y/\sqrt{y}$)
will also be significant for these observables in the kinematics of
actual experiments. The systematic calculation of such corrections
is presently not within reach, but it would be an exciting and useful interdisciplinary
program.


\section*{Acknowledgements}
We thank Dr. St\'ephane Peign\'e for his interest in this research.
The work of ADL and SM is supported in part by the Agence Nationale
de la Recherche under the project ANR-16-CE31-0019.
The  work  of  AHM  is  supported in part by
the U.S. Department of Energy Grant
DE-FG02-92ER40699.


\appendix

\section{A few useful integrals\label{app:integral}}

The calculations presented in the body of this paper
require to perform a few integrals.
Let us introduce the following notations:
\be
  I_{k}(A)\equiv\int_1^\infty\frac{dt}{t^{2}}
  \left[1-\sum_{i=0}^{k-1}\frac{(At)^i}{i!}e^{-At}\right]
\quad\text{and}\quad
  I'_{k}(A)\equiv\int_1^\infty\frac{dt}{t^{2}}\ln t\,
  \left[1-\sum_{i=0}^{k-1}\frac{(At)^i}{i!}e^{-At}\right].
  \ee
We will study the following particular cases:  
\begin{align}
  I_1(A)&\equiv\int_1^\infty\frac{dt}{t^2}\left(1-e^{-At}\right)
  &I_1'(A)&\equiv \int_1^\infty\frac{dt}{t^2}\ln t\left(1-e^{-At}\right)\\
  I_{2}(A)&\equiv\int_1^\infty\frac{dt}{t^{2}}
  \left[1-(1+At)e^{-At}\right]
  &I_{2}'(A)
  &\equiv\int_1^\infty\frac{dt}{t^{2}}\ln t\,
  \left[1-(1+At)e^{-At}\right]
\end{align}
Although these integrals have exact expressions in terms of
special functions, only the small-$A$
limits will be relevant for our purpose.

These integrals can be deduced from
more general ones:
\be
\begin{split}
  I_{1,\varepsilon}(A)&\equiv\int_1^\infty\frac{dt}{t^{2-\varepsilon}}
\left(1-e^{-At}\right)
=\frac{1}{1-\varepsilon}\left[1-e^{-A}
+\Gamma(\varepsilon,A)A^{1-\varepsilon}\right],\\
I_{2,\varepsilon}(A)&\equiv\int_1^\infty\frac{dt}{t^{2-\varepsilon}}
\left[1-(1+At)e^{-At}\right]
=\frac{1}{1-\varepsilon}\left[1-e^{-A}
+\varepsilon\Gamma(\varepsilon,A)A^{1-\varepsilon}\right],
\end{split}
\ee
where $\Gamma$ is the incomplete Gamma function:
\be
\Gamma(\varepsilon,A)\equiv\int_A^\infty
d\bar t\,{\bar t}^{\varepsilon-1}e^{-\bar t}.
\ee
We want to calculate the leading term in the small-$A$ limit of
the expansion of these integrals to order $\varepsilon$.
To this aim, we write
\be
\begin{split}
\Gamma(\varepsilon,A)&=\frac{1}{\varepsilon}
\left[\Gamma(1+\varepsilon)-A^\varepsilon+{\cal O}(A)\right]\\
&=\ln\frac{1}{A}-\gamma_E+
{\cal O}(A)+\frac{\varepsilon}{2}
\left[\psi'(1)-\ln^2\frac{1}{A}+{\cal O}(A)\right]+\cdots
\end{split}
\ee
Thus
\be
I_1(A)=A\,\ln\frac{1}{A}+{\cal O}(A)
\quad\text{and}\quad
I_1'(A)=\frac{A}{2}\ln\frac{1}{A}
\left[\ln\frac{1}{A}+2(\psi(1)+1)\right]
+{\cal O}(A).
\label{eq:I1}
\ee
Note that the factor $A$ is dimensional, while $1/A$ in the argument
of the logarithm is just the size
of the relevant integration region, which extends to $\sim 1/A$.
Actually, the integral could also be performed by simply
restricting the integration region to $[1,\kappa/A]$, where
$\kappa\sim{\cal O}(1)$, and expanding
the exponential: The overall normalization of the leading term
for $A\ll 1$ 
would be identical to the one found from the exact calculation.
The details of how the integration region is effectively cut off
in the integrand do not matter.

As for $I_{2}$ and $I_{2}'$, 
we expand $I_{2,\varepsilon}$ to order $\varepsilon$. We arrive at
the following exact expressions, together with their expansion at lowest
order for $A\ll 1$:
\be
I_{2}(A)=1-e^{-A}\simeq A[1+{\cal O}(A)]
\quad\text{and}\quad
I_{2}'(A)=A\,\Gamma(0,A)\simeq A\,\ln\frac{1}{A}+{\cal O}(A).
\label{eq:I2}
\ee
Note that $I_1$ and $I_{2}'$ are identical for small $A$.
However, $I_2'$ is not a logarithmic integral, and therefore,
the overall constant factor of the leading term in the limit $A\ll 1$
of interest here is sensitive to the detailed form of the integrand.


\section{Numerical calculations in an alternative model\label{app:num}}

The branching random walk model defined and studied in Sec.~\ref{sec:numerics} was
actually a particular discretization, in space and time, of a branching Brownian
motion.
The FKPP equation~\cite{f,kpp}, that is solved e.g. by the probability that there is at least
a particle to the right of some predefined $X$, when the diffusion constant of
the Brownian process is set to $D$, reads
\be
\partial_y T_1=D\,\partial_x^2 T_1+T_1(1-T_1).
\label{eq:FKPP}
\ee
Equation~(\ref{eq:equivBK}) for $S=1-T_1$
is a particular lattice discretization of~(\ref{eq:FKPP}),
with $D=\frac12$.

In this Appendix, we shall study an alternative discretization, namely
\be
T_1(y+\delta y,x)- T_1(y,x)=
\frac{\delta y}{\delta x^2} \left[T_1(y,x+\delta x)-2T_1(y,x)+T_1(y,x-\delta x)\right]
+\delta y\, T_1(y,x)[1-T_1(y,x)].
\ee
We would recover Eq.~(\ref{eq:FKPP}) with $D=1$
in the joint limit $\delta y,\delta x\rightarrow 0$.

The underlying branching random walks are actually quite different in the two
discretization schemes. In the present scheme,
a particle on a given site has a probability to move, either left or right,
of order $\delta y$, which is taken small.
In the previous scheme instead, it had a probability $1-\delta y$ to move.

In this new scheme, the evolution equation for $G$ reads, for $y>y_1$,
\begin{multline}
  G(y+\delta y,x;y_1)-G(y,x;y_1)=
  \frac{\delta y}{\delta x^2} \left[G(y,x+\delta x;y_1)
    -2G(y,x;y_1)+G(y,x-\delta x,y;y_1)\right]\\
  +\delta y \,G(y,x;y_1)[1-2T_1(y,x)],
	\label{eq:G_discrete}
\end{multline}
and the initial condition is the same as in the previous model, see Eq.~(\ref{eq:initGmodel}).

We have set $\delta y=0.02$ and $\delta x=0.25$, which leads to the
following value of the constants entering the asymptotic quantities
of interest:
\be
\gamma_0 = 1.0120279\cdots,\quad
\chi'(\gamma_0) = 1.9659159\cdots,
\quad \chi''(\gamma_0) = 1.9065278\cdots
\label{eq:par2}
\ee
These parameters are quite different from those of the first model;
Compare to Eq.~(\ref{eq:par1}).
We have also chosen a different initial condition: $T_1(y=0,x)=1-\exp(e^{-4\gamma_0 x})$.

Our implementation of these evolution equations is independent
of the one of the first model,
and the very numerical methods used are different.
As for the model exposed here, we evolve
$\ln T_1$ and $\ln G$ instead of $T_1$ and $G$ directly, at variance with the
first model, in order to make
sure that we treat accurately-enough the crucial region in which
these functions take very small values (see e.g. \cite{Mueller:2014gpa}
for a description of such a numerical method).

\begin{figure}
  \begin{center}
    \includegraphics[width=.9\textwidth]{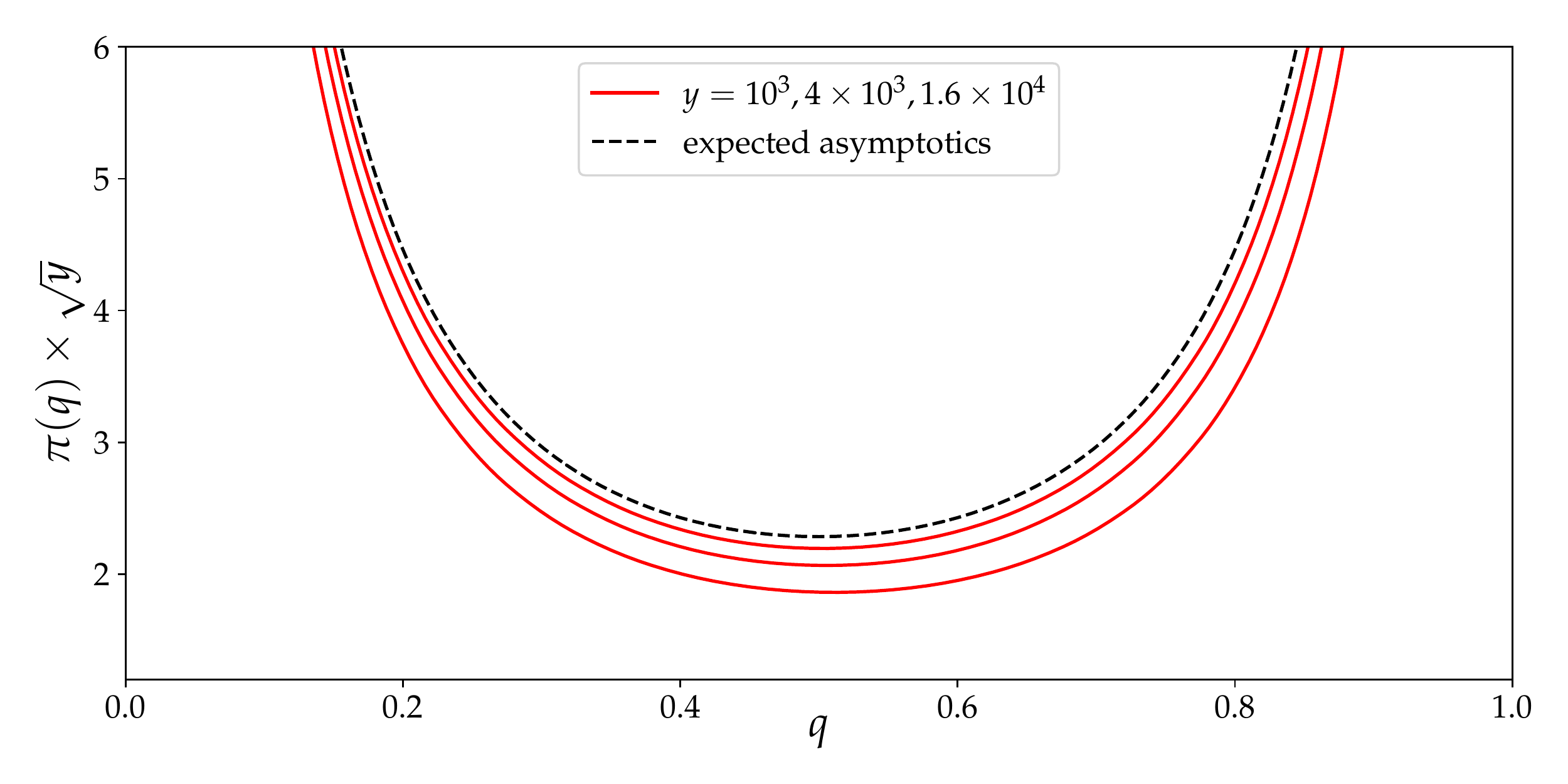}
  \end{center}
  \caption{{\small Distribution of the overlaps for different values of $y$
    (set of full lines; $y\in\{10^3,4\times 10^3,1.6\times 10^4\}$ increasing
    from bottom to top) and $X-X_y$ set to $\sqrt{2}\,y^{1/4}$,
    together with the expected asymptotics given in Eq.~(\ref{eq:overlap})
    with the  constants~(\ref{eq:par2})
    (dashed line).}\label{fig:overlap_B}}
\end{figure}

The overlaps $\pi_y(q)$, shown in Fig.~\ref{fig:overlap_B} for
$y\in \{10^3, 4\times 10^3,1.6\times 10^4\}$, are compatible with a convergence
to the expected asymptotics~(\ref{eq:overlap}) -- although it is more
difficult to judge than for the first numerical model: The maximum value
of $y$ for which we have been able to perform the calculation is
more than one order of magnitude smaller than in the case of the latter.



\end{document}